\documentclass[%
 reprint, showkeys,
 amsmath,amssymb,
 aps,
]{revtex4-2}

\usepackage{graphicx}
\usepackage{pgfplots}
\usepackage{dcolumn}
\usepackage{bm}
\usepackage{hyperref}

\begin{document}

\preprint{APS/123-QED}

\title{Stochastic drift in discrete waves of non-locally interacting-particles}

\author{Andrei Sontag}%
\email{ams284@bath.ac.uk}
\author{Tim Rogers}%
\author{Christian A. Yates}%
\affiliation{%
 Centre for Mathematical Biology,\\ Department of Mathematical Sciences, University of Bath, Bath, BA2 7AY, UK
}%
\date{ }




\date{\today}

\begin{abstract}

In this paper, we investigate a generalised model of $N$ particles undergoing second-order non-local interactions on a lattice. Our results have applications across many research areas, including the modelling of migration, information dynamics and Muller's ratchet -- the irreversible accumulation of deleterious mutations in an evolving population. Strikingly, numerical simulations of the model are observed to deviate significantly from its mean-field approximation even for large population sizes.
We show that the disagreement between deterministic and stochastic solutions stems from finite-size effects that change the propagation speed and cause the position of the wave to fluctuate. These effects are shown to decay anomalously as $(\log N)^{-2}$ and $(\log N)^{-3}$, respectively -- much slower than the usual $N^{-1/2}$ factor. Our results suggest that the accumulation of deleterious mutations in a Muller’s ratchet and the loss of awareness in a population may occur much faster than predicted by the corresponding deterministic models. The general applicability of our model suggests that this unexpected scaling could be important in a wide range of real-world applications.
\end{abstract}

\maketitle

\section{Introduction}\label{sec:introduction}

Models of interacting particles on lattices are ubiquitous. Their applications range from spin interactions~\cite{PhysRev.87.410}, to voter models~\cite{Holley_1975,Liggett1999}, and epidemics~\cite{RHODES1996125,RHODES1997101,GRASSBERGER1983157}, to cite but a few. Moreover, lattice models are particularly convenient to describe the movement of animals, microorganisms and cells~\cite{Codling_2008,taylor2015dab}. Given their importance, research has also been devoted to the study interacting-particle lattice models with no particular application in mind~\cite{Dickman_1989,PhysRevE.83.051922,Penrose_2008}.

Due to their complexity, the analysis of these stochastic systems is often non-trivial. Numerical simulations and deterministic approximations are common approaches to gain insight into complex interacting-particle systems. Stochastic simulations allow for a higher degree of realism and precision at the expense of increased computational cost and limited generalisation and understanding of the results obtained. In contrast, an easy-to-extend analytical result is often sought by the introduction of moment-closures and the hydrodynamic limit, which provide more accessible deterministic equations. This approach is frequently dubbed a `mean-field approximation'~\cite{2014Toral}.

The power of mean-field approximations, to reduce complexity of stochastic systems by neglecting noise effects, makes this analytical device a cornerstone in mathematical biology modelling. For many stochastic systems in biology, the effects of finite size populations are frequently small, decaying as the square root of the population size. Thus, the use of mean-field approximations provides a simplified — but still complex — description of the systems' dynamics that can be analysed by dynamical systems' theory without appreciable loss in accuracy and predictability~\cite{Bick_2020}.

Although these deterministic approximations usually describe well systems with many particles -- or the averages of numerous simulations -- conclusions drawn from them should be evaluated cautiously. Stochasticity causes an assortment of fascinating dynamics not predicted by the mean-field approximation. For instance, noise can amplify and maintain seasonality in epidemics~\cite{Alonso_2006}, induce bistability in collective behaviour~\cite{PhysRevLett.112.038101}, or even completely reverse the direction of deterministic selection~\cite{Constable_2016}.


In this article, we investigate a generalised model of $N$ particles undergoing second-order non-local interactions on a lattice. The basic model we build upon was first introduced as an `awareness spread' model in~\cite{Funk2009}, concomitantly with disease dynamics, to investigate the effect of behavioural reactions on epidemics. In this model, subpopulations are characterised by an index $i\in\mathbb{N}_0$ that gives the quality of the information held by an individual. The smaller the value of $i$, the better the information quality. The awareness dynamics are simple: information can be transmitted from a better informed individual to another, losing quality in the process, which increases the index of the receiving individual by $1$ after transmission; and individuals spontaneously lose awareness if their information is not refreshed, also increasing the index by $1$. The `awareness spread' model exhibits stochastic wavefront solutions. This and the discrete nature of the individual's `information quality' index are the two distinguishing features of this model that attract our attention. For this reason, our model will consist of a fixed-size population embedded in a generalised discrete space that can be either abstract -- such as the `information quality space' in the awareness model, and the `niche space' in genetics -- or real space. Thus, our generalised model has applications in many areas of research, ranging from ecology to collective motion, cell division and motility, and evolution. In particular, the original model can be understood as a Muller's ratchet. Named after Hermann Joseph Muller, Muller's ratchet refers to an irreversible process in evolutionary genetics by which asexual populations accumulate deleterious mutations as a consequence of genetic drift and mutation, amplified by a small population size~\cite{10.1093/genetics/78.2.737}.

\begin{figure}[t]
\includegraphics[width=8.6cm]{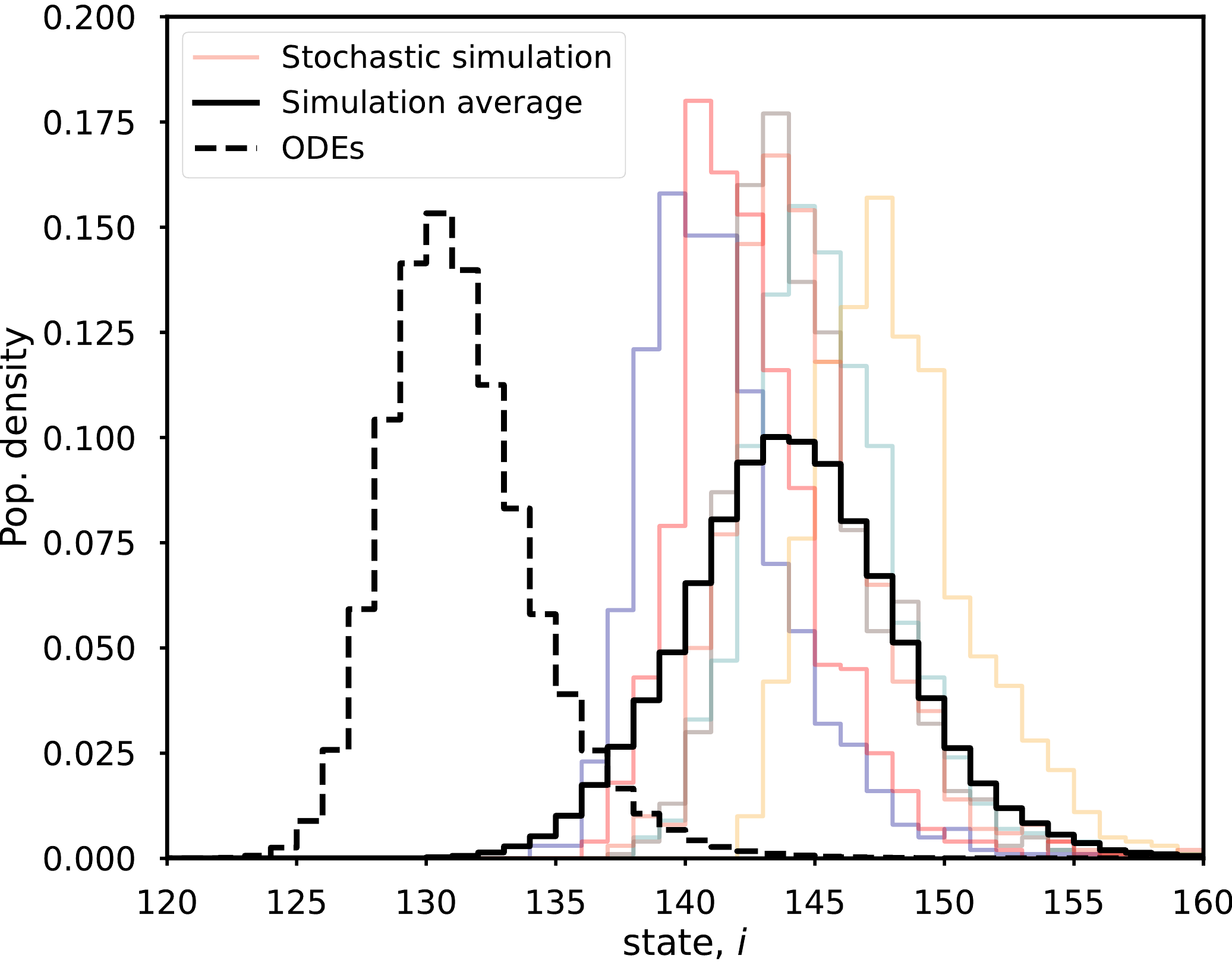}
\caption{\label{fig:example_sim} The stochastic waves (colours) propagate faster than the deterministic solution (dashed). The position of the wave at a given time fluctuates from simulation to simulation, flattening the average curve (solid black line). The parameters used to generate the figure are: $N=10^3$, $t = 400$, $\alpha = 0.5$, $\beta = 0$, $\lambda_1 = 1$ and $\lambda_2 = 0$ (see Sec.~\ref{sec:model} for the model description). Initial condition: $n_{50}(0) = N$.}
\end{figure}

As we will see, numerical simulations have shown substantial differences between the stochastic waves and the corresponding mean-field approximation in the two cases agreement could be expected, i.e., for large number of particles and for the average of many simulations. As we will demonstrate later through numerical simulations and analytically, applying the cut-off method developed by~\citet{Brunet_1997}, this disagreement is caused by the drift and fluctuation of the stochastic wavefront, effects that decay respectively as $(\log N)^{-2}$ and $(\log N)^{-3}$ with the population size, $N$, much slower than the usual $N^{-1/2}$ factor. Readers familiar with the work of Kurtz~\cite{Baxendale2010} or Van Kampen~\cite{VANKAMPEN2007244} may find this difference surprising. Indeed, it is a proven theorem that the fluctuations in the number of individuals at index $i$ at time $t$ will decay as $1/\sqrt{N}$~\cite{KURTZ1978223}. The important thing to note here is that the question of the speed of a wave does not concern events taking place up to a certain point in time, but rather the time elapsed until a certain event takes place. This small-seeming semantic difference has a fundamental impact on the large-$N$ scaling laws, as seen here and elsewhere (e.g.~\cite{Parsons2018}).

Fig.~\ref{fig:example_sim} gives an example of this discrepancy. Although the deterministic solution captures well the shape of individual realisations of the stochastic wave, there is a significant difference in wave speed between them. Furthermore, by default, the deterministic solution does not reproduce the fluctuations in the wavefront position that lead to a flattened average curve. A consequence of these effects is a much faster accumulation of deleterious mutations in the ratchet model, and similarly the loss of awareness in a population, than predicted by the deterministic limit.

The anomalous scaling observed in the convergence of the speed of propagation and diffusion of the wavefront position has been studied in the literature in the context of partial differential equations (PDEs) with a cut-off~\cite{Dumortier_2007,Dumortier_2014}, branching Brownian motion~\cite{Brunet_2006,Brard2010}, and diffusing particle systems that interact locally~\cite{BREUER1994259,Kessler1998,Panja_2002,PhysRevE.59.3893}.
The novelty in our work comes from analysing a more general model of interacting-particles that perform a biased random walk on a semi-infinite lattice and non-local interactions with rate which depends on the distance between particles. We show that this model can be analysed in terms of a cumulative variable for which the dynamics, under certain conditions, can be approximated by discrete KPP-like equations. In this case, the equations depend only on the rate at which particles jump large distances.

The article is organised as follows. In Sec.~\ref{sec:model}, we generalise the model introduced in~\cite{Funk2009}. We introduce a moment-closure to obtain a system of ODEs that approximate the dynamics of the system in Sec.~\ref{sec:moment_closure}, and take the continuous limit in Sec.~\ref{sec:continuous_limit} to obtain an Integro-PDE description. For particular cases, a useful change of variables yields a generalised Kolmogorov-Petrovsky-Piscounov (KPP) equation~\cite{kpp_equation}. Results are presented in Sec.~\ref{sec:speed}. In particular, we obtain an analytical correction for the stochastic wavefront's speed and compare this prediction to results from simulations. In Sec.~\ref{sec:discussion} we discuss and give insight into other classes of models that we expect to present similar disagreement between the stochastic and the mean-field solutions. 

\section{Model}\label{sec:model}

The model we study here is a generalisation of an awareness spread model introduced in~\cite{Funk2009}. The authors couple the awareness dynamics to a SIR model to study the interplay of disease outbreaks and behavioural responses -- prompted by awareness transmission in the population. In~\cite{Sontag_2022}, the model was extended to include contrasting behaviours regarding the transmission of information, with individuals seeking information of better or worse quality depending on the subpopulation they belong to. 

Our system consists of a population of fixed size, $N$, in which each individual $X$ has an associated index $i\in\mathbb{N}_0$ indicating the state in which the individual is. In the awareness spread model, this corresponds to the quality of the information that the individual has. 

The governing dynamics of our generalised model are:
\begin{enumerate}
    \item \textit{Pairwise copying}, 
    \begin{eqnarray}
    X_i + X_j &&\xrightarrow{f(i-j)} 2X_i.
    \end{eqnarray}
    \item \textit{Individual updates}, 
    \begin{eqnarray}
    X_i \mathrel{\mathop{\rightleftarrows}^{\lambda_1}_{\lambda_2}} X_{i+1}.
    \end{eqnarray}
\end{enumerate}

We assume that the pairwise copying occurs with a rate given by a function of the difference in states between the two interacting individuals, $f:\mathbb{R}\to\mathbb{R}^+_0$. In the definition above, $f(z)$ is a function with inputs in on $\mathbb{Z}$, but later on its domain will be extended to $\mathbb{R}$. 
This modelling choice of $f$ is motivated by competition kernels commonly adopted in the ecological and evolutionary literature~\cite{Rogers_2012,Fuentes_2003} which depend on the species' distance in niche space. In contrast to these models, the function $f$ is not necessarily even; $f(x) \neq f(-x)$ in general. This uneveness is a necessary condition to observe the propagation of stochastic wavefronts that disagree with its mean-field approximation (see Appendix~\ref{apx:symmetric} for details). 

The second interaction can be interpreted as a biased random walk on a lattice. In an awareness model, the onwards reaction accounts for a memory loss process in the population (individuals will slowly lose awareness with rate $\lambda_1$ if their information is not refreshed), whereas the backwards reaction describes the finding of better quality information, through research or similar means, with rate $\lambda_2$.

In the original model,~\cite{Funk2009}, information transmission occurs solely from better informed to worse informed individuals, $f(i-j) = \alpha\cdot\theta(j-i)$, where $\theta(z)$ is the Heaviside function, and the spontaneous acquisition of better information is absent, $\lambda_2=0$. In that case, if we index by $i = 0$ the source of information, then the subscript $i$ of an individual indicates how many times the acquired information has faded. Once all sources of the best quality of information have faded, there is no way to generate new information. Hence, information is gradually lost and eventually disappears if not refreshed. This model can also be interpreted as a Muller's ratchet, where the index $i$ counts the number of deleterious mutations that an individual has accumulated. Competitors with fewer mutations outcompete their rivals and reproduce. The offspring then replaces the losing competitor in the population to keep its size constant.

Besides the Muller's ratchet interpretation, our model has several applications in distinct areas of research for more general interaction functions, $f$. As an ecological model, it can be understood as a species competing for resources and performing a biased random walk on a lattice. As a continuous-time evolutionary model, we may take the index $i$ as the individual's position in fitness space, whilst $\lambda_1$ and $\lambda_2$ give the rates of mutation to worse and better fitnesses and the function $f(z)$ is the competition kernel between different strains. As a cell cycle-synchronisation process, $\lambda_1$ gives the rate at which cells progress to the next phase of their cycle, $\lambda_2 = 0$, and the function $f$ synchronises their phase.

To facilitate the analysis of this stochastic model, in the next section, we introduce its corresponding master equation and derive a system of ODEs based on a moment-closure approximation. In Sec.~\ref{sec:continuous_limit} we take the hydrodynamic limit to obtain a continuous description of the model, and obtain results on the dependency of the speed of propagation of the wavefront on the choice of functions $f(z)$ that motivate the rest of our work.


\section{Moment-closure approximation}\label{sec:moment_closure}
Using master equations, we can describe the evolution of probabilities for Markov processes that evolve over time from one state to another~\cite{2014Toral}. 
For this system, the master equation reads
\begin{widetext}
\begin{eqnarray}
\frac{\partial }{\partial t}p(n_0,n_1,...;t) =&&\ \sum_{i=0}^\infty\sum_{j=0}^\infty \frac{f(i-j)}{N}(n_i-1)(n_j+1) \cdot p(n_0,n_1,...,n_i-1,\dots,n_{j} + 1, \dots; t)\nonumber\\
    &&+ \sum_{i=0}^\infty (n_i + 1)\lambda_1 \cdot p(n_0,n_1,...,n_i+1,n_{i+1}-1,\dots; t)\nonumber\\
    &&+ \sum_{i=0}^\infty (n_i + 1)\lambda_2 \cdot p(n_0,n_1,...,n_{i-1}-1,n_i+1,\dots; t)\nonumber\\
    &&- \left(\sum_{i=0}^\infty \sum_{j=0}^\infty \frac{f(i-j)}{N}n_in_j + \sum_{i=0}^\infty (\lambda_1 + \lambda_2)n_i \right)\cdot p(n_0, n_1,\dots; t),\label{eq:master_equation}
\end{eqnarray}
\end{widetext}
where $p(n_0,n_1,\dots; t)$ is the probability that, at time $t$, there are $n_0$ individuals in state $0$, $n_1$ in state $1$, and so on. Furthermore, we assume $p(n_0,n_1,\dots,n_i,\dots; t) = 0$ if $n_i < 0$ for any $i$. Note that $\sum_{i=0}^\infty n_i = N$, so only a finite number of $n_i$'s will be non-zero at any given time. 

Although the master equation gives a more precise description of the time evolution of the system, it is often not possible to solve analytically. A common approach used to obtain insight about the dynamics of the model is the mean-field approximation. Multiplying Eq.~\eqref{eq:master_equation} by $n_k$ and summing over the state space gives
\begin{eqnarray*}
    \frac{d\langle n_k\rangle}{dt} =&& \sum_{i} \left\langle n_k n_i \frac{f(k-i)}{N} \right\rangle - \sum_{i} \left\langle n_k n_i \frac{f(i-k)}{N} \right\rangle\nonumber\\
    &&+ \lambda_1 \langle n_{k-1}\rangle-(\lambda_1+\lambda_2) \langle n_k\rangle + \lambda_2 \langle n_{k+1}\rangle.
\end{eqnarray*}

Despite being an exact equation for the mean population size in each compartment, this system of equations is not closed, as it depends on second-order moments. To close the system, we assume the moment closure $\langle n_in_j\rangle = \langle n_i \rangle \langle n_j\rangle$ for any $i$ and $j$, including $i = j$, obtaining the following mean-field approximation,
\begin{eqnarray}\label{eq:moment_closure}
    \frac{d\langle n_k\rangle}{dt} =&& \lambda_1\langle n_{k-1}\rangle-(\lambda_1+\lambda_2)\langle n_k\rangle + \lambda_2 \langle n_{k+1} \rangle\nonumber\\
    &&+ \frac{1}{N}\sum_{i} \langle n_k\rangle\langle n_i\rangle \left(f(k-i) -  f(i-k)\right).\label{eq:mf_discrete}
\end{eqnarray}

This equation is already useful for some calculations we will develop later, however, in order to take advantage of the literature on PDEs, and to obtain results that motivate our work, we need to take the continuous limit.

\section{The continuous limit}\label{sec:continuous_limit}
One may recognize Eq.~\eqref{eq:moment_closure} as an on-lattice discretisation of a drift-diffusion jump process with a reaction term (see, e.g.,~\cite{Stevens_1997,baker2009fmm}). In these models, it is common to introduce the variable $u(x,t)$ as the continuous limit of the vector $\mathbf{n}(t) = [n_0(t)/N, n_1(t)/N, \dots]$. This yields the corresponding Integro-PDE description of the system (for details of the derivation of the continuum limit see, for example,~\cite{taylor2015dab})
\begin{align}
    \frac{\partial u}{\partial t} =&\ (\lambda_2-\lambda_1) h \frac{\partial u}{\partial x} + \frac{(\lambda_1+\lambda_2) h^2}{2}\frac{\partial^2 u}{\partial x^2}\label{eq:continuous_limit}\\
    &+ \int_0^{\infty}u(x,t) u(y,t)\left[ f(x-y) - f(y-x)\right]\text{d}y,\nonumber
\end{align}
where $h$ is the lattice cell size, and $u(x,t)$ is the density of the population in the interval $[x,x+\text{d}x)$ at time $t$. 

We use boundary conditions $u(\infty) = u_x(\infty) = 0$, and
\begin{align}
	0 =& (\lambda_1 - \lambda_2)h\hspace{0.05cm} u(0,t) + \frac{(\lambda_1+\lambda_2)h^2}{2}\frac{\partial u(0,t)}{\partial x},
\end{align}
which enforce conservation of mass. Because, $u(x,t) \geq 0, \forall x$, and $\int_0^\infty u(y,t)\text{d}y = 1$, $u(x,t)$ can be interpreted as a probability density function. This, and the integral term, suggest the change of variable $U(x,t) = \int_0^x u(y,t)\text{d}y$, the cumulative sum of the number of individuals with position $y < x$, in order to simplify Eq.~\eqref{eq:continuous_limit}. Note that $U(0) = 0$ and $\lim_{x\to\infty}U(x) = 1$. Without loss of generality, we will take $h=1$ for the rest of this paper, unless stated otherwise.

Integrating Eq.~\eqref{eq:continuous_limit} on the interval $(0,x)$ yields
\begin{eqnarray*}
    \frac{\partial U}{\partial t} =&&\ (\lambda_2-\lambda_1)  \frac{\partial U}{\partial x} + \frac{(\lambda_1+\lambda_2)}{2}\frac{\partial^2 U}{\partial x^2}\label{eq:continuous_limit2}\\
    &&+ \int_0^x u(y,t)\int_0^{\infty} u(z,t)\left[ f(y-z) - f(z-y)\right]\text{d}z\text{d}y.\nonumber
\end{eqnarray*}

This is now an equation for the new variable $U(x,t)$. We can work further on the double integral with integration by parts,
\begin{eqnarray*}
    &&\int_0^{\infty} u(z,t)\left[ f(y-z) - f(z-y)\right]\text{d}z\\
    &&\quad\quad =\ [U(z,t)(f(y-z)-f(z-y))]^\infty_0\\
    &&\quad\quad\quad\ - \int_0^\infty U(z,t)(f'(y-z) - f'(z-y))\text{d}z,\\
    &&\quad\quad =\ f^- - f^+ - \int_0^\infty U(z,t)(f'(y-z) - f'(z-y))\text{d}z,
\end{eqnarray*}
\noindent where $f^\pm = \lim_{z\to\infty}f(\pm z)$, and the derivatives $f'(z)$ must be understood in the weak sense. Substituting back yields
\begin{widetext}
\begin{eqnarray*}
    \frac{\partial U}{\partial t} =&&\ (\lambda_2-\lambda_1) \frac{\partial U}{\partial x} + \frac{(\lambda_1+\lambda_2)}{2}\frac{\partial^2 U}{\partial x^2}+  U(f^- - f^+) - \int_0^x u(y,t)\int_0^\infty U(z,t)(f'(y-z) - f'(z-y))\text{d}z\text{d}y.
\end{eqnarray*}
\end{widetext}

This is a reaction-advection-diffusion equation for the variable $U(x,t)$ which may have wavefront solutions depending on the choices of $f(z)$. Note that the integral contribution is of order higher than linear. Provided that this contribution is sufficiently small compared to the linear term, we can linearise the equation,
\begin{equation*}
    \frac{\partial U}{\partial t} =\ (\lambda_2-\lambda_1)  \frac{\partial U}{\partial x} + \frac{(\lambda_1+\lambda_2)}{2}\frac{\partial^2 U}{\partial x^2}+  U(f^- - f^+),
\end{equation*}
and look for wavefront solutions, with speed $v$, of the form $U(x,t) = U(x+vt) = Ce^{\gamma(x+vt)}$, where $\gamma$ is the exponential growth rate at the front of the wave. Substituting results in an expression relating $v$ and $\gamma$,
\begin{equation}\label{eq:speed_on_f}
    \gamma (v + \lambda_1 - \lambda_2) = \frac{(\lambda_1+\lambda_2)}{2}\gamma^2 + (f^- - f^+).
\end{equation}

As for any KPP-equation, Eq.~\eqref{eq:speed_on_f} gives a family of pairs $(v,\gamma)$. The direction of propagation is determined by the sign of $v$, which depends on parameter values. The critical speed is given by 
\begin{equation*}
	v_c = (\lambda_2-\lambda_1) + 2\hspace{0.05cm}\text{sgn}(f^- - f^+)\sqrt{\tfrac{(\lambda_1+\lambda_2)}{2}|f^- - f^+|},
\end{equation*}
which agrees with previous results on generalised-KPP equations~\cite{Kollr2016}. If $f^--f^+ > 0$, then the wave propagates with speed $v \geq v_c$, whilst $v \leq v_c$ if $f^--f^+ < 0$. Notably, the speed of propagation depends on $f(z)$ only through the difference $f^{-} - f^{+}$, i.e., the limits $\lim_{z\to\pm\infty} f(z)$. 


In Appendix~\ref{apx:polynomial} we analyse polynomial functions of the distance between interacting states and show that, for a modification of the original model~\cite{Funk2009}, no wavefronts are obtained in this case. In this paper, we focus on choices of $f(z)$ that lead to wavefronts. Consequently, the dependency of the wave speed only through the limits $\lim_{z\to\pm\infty} f(z)$ motivates us to investigate functions of the form $f(z) = \alpha \cdot \theta(-z) + \beta \cdot \theta(z)$ for any choice of real $\alpha$ and $\beta$ as a proxy for more general functions as a means to obtain analytical results that can be easily extended. 
In the next section, we develop the results obtained by this choice of $f(z)$. In particular, we observe a discrepancy between the deterministic solutions and the average of many simulations, as well as for large population sizes -- limits in which the mean-field approximation is often assumed to be a good description of a stochastic system. In Sec.~\ref{sec:discussion} we discuss what other classes of functions are expected to present similar disagreement with the mean-field solution.


\section{Results}\label{sec:speed}

For our choice of $f(z) = \alpha \cdot \theta(-z) + \beta \cdot \theta(z)$, the discrete system of equations in the mean-field approximation is 
\begin{eqnarray}\label{eq:mf_ode_first}
    \frac{d \langle n_k\rangle}{dt} =&& \lambda_1  \langle n_{k-1}\rangle-(\lambda_1+\lambda_2)  \langle n_k\rangle + \lambda_2  \langle n_{k+1}\rangle \nonumber\\
    &&+ \frac{\alpha}{N}\sum_{i>k} \langle n_k\rangle \langle n_i\rangle - \frac{\beta}{N}\sum_{i < k} \langle n_k\rangle \langle n_i\rangle,
\end{eqnarray}
or, for the new variable $U_k(t) = \frac{1}{N}\sum_{i=0}^k \langle n_i(t)\rangle$,
\begin{eqnarray}\label{eq:mf_ode}
    \frac{d U_k}{dt} =&&\ \lambda_1  U_{k-1}-(\lambda_1+\lambda_2)  U_k + \lambda_2  U_{k+1}\nonumber\\
    &&+\ (\alpha-\beta)U_k(1-U_k).
\end{eqnarray}
Hence, in the mean-field approximation, the system with non-local jumps to earlier (left) or later (right) states with rates $\alpha$ and $\beta$, respectively, is equivalent to the system with non-local jumps to the left with rate $\alpha - \beta$ if $\alpha > \beta$, or to the right with rate $\beta - \alpha$ if $\beta > \alpha$. Note that the continuous limit of Eq.~\eqref{eq:mf_ode} yields an advection-reaction-diffusion equation with logistic growth. In the case $\lambda_1 = \lambda_2 = \lambda$, the advection term vanishes, and we obtain a Kolmogorov-Petrovsky-Piscounov (KPP) equation~\cite{kpp_equation},
\begin{equation}\label{eq:reac-diff-adv}
    \frac{\partial U}{\partial t} =\ \lambda\frac{\partial^2 U}{\partial x^2}+  (\alpha-\beta)U(1-U).
\end{equation}

KPP-equations are one of the most fundamental models in mathematical biology~\cite{2002Murray}, commonly used to describe population dynamics in space and time. A well-known feature of this class of equations is the existence of a family of wavefront solutions defined by $\gamma > 0$ with speed $v(\gamma) = \lambda\gamma + \frac{\zeta}{\gamma}$, where $\zeta = \alpha - \beta > 0$. For general, non-negative, initial conditions, the wave propagates with speed $v \geq v_c = 2\sqrt{\lambda\zeta}$, achieving the critical speed $v_c$ for well-behaved initial conditions. Wavefronts with speed $v < v_c$ require $\gamma$ to be complex, and are thus not physical, as convergence to these solutions is not achieved by any realistic initial condition. We expect our discrete system of equations to exhibit similar behaviour. 


\subsection{Numerical Results}\label{sec:numerical_results}

In this section, we investigate the behaviour of the stochastic wave numerically for our choice of $f(z)$, and compare it to the solution of the deterministic system of equations, Eq.~\eqref{eq:mf_ode_first}. Without loss of generality, we will assume $\alpha > \beta$ for the rest of this paper, unless stated otherwise.  The case $\alpha = \beta$ is a special case of a symmetric jump, and hence, agrees well with the moment-closure approximation (see Appendix~\ref{apx:symmetric}). All conclusions obtained for $\alpha > \beta$ hold directly for $\beta > \alpha$ with $\lambda_1$ and $\lambda_2$ interchanged, and reflected state direction.

Fig.~\ref{fig:example_sim} compares a stochastic simulation of the system and the solution of the discrete system of equations for $\lambda_2 = 0$. This case is of special interest as it corresponds to the original awareness spread model~\cite{Funk2009} and to a Muller's ratchet. There is significant disagreement between the two solutions, with the stochastic wavefront propagating faster than its deterministic counterpart. This effect suggests that the accumulation of deleterious mutations in a population, or the loss of awareness, are processes that occur faster than predicted by the deterministic model. This might have critical consequences for the survival of species, or the outcome of an epidemic, in which the spread of awareness plays a vital role in eliciting behavioural reactions and increasing adoption of non-pharmaceutical interventions.

\begin{figure}[t]
\includegraphics[width=8.6cm]{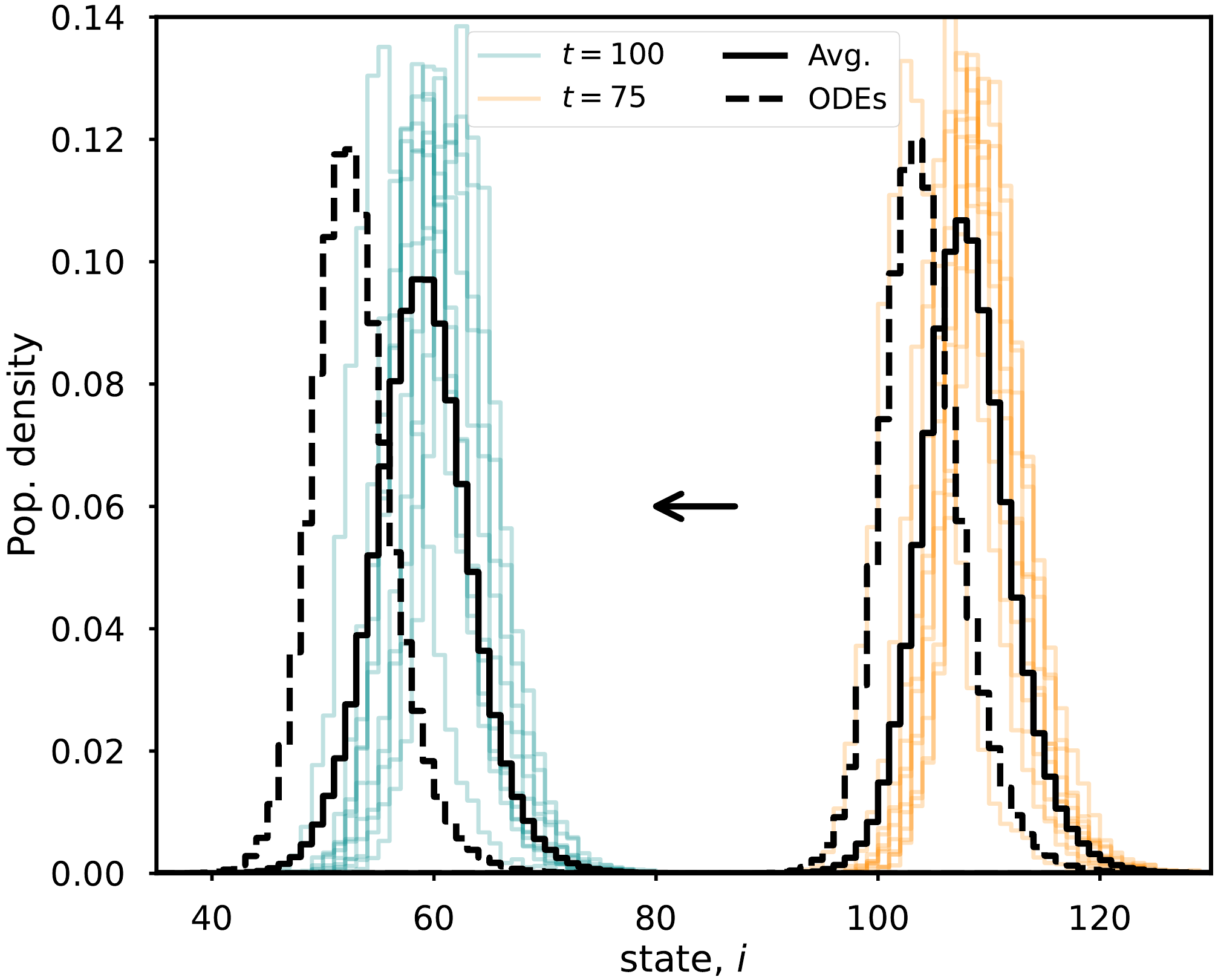}
\caption{\label{fig:same_lambda} Comparison between the stochastic and the deterministic systems for $\lambda_1 = \lambda_2$ at different instants of time, $t$. The solution of the ODE system, Eq.~\eqref{eq:mf_ode}, (dashed) is faster than the stochastic waves (colours). The flattening of the simulations' average (solid black line) is caused by fluctuations in the position of the stochastic wave from realisation to realisation. Parameters: $N=10^4$, $\alpha = 1$, $\beta = 0$, $\lambda_1 = \lambda_2 = 1$. Initial condition: $n_{250}(0) = N$.}
\end{figure}

Another case of interest is $\lambda_1 = \lambda_2 = \lambda$. In this setting, the advection term vanishes and the wavefront propagates to lower states ($\alpha > \beta$). This movement is driven by individuals migrating leftwards as a result of the spontaneous update to a previous state and the pairwise copying interactions.
Fig.~\ref{fig:same_lambda} compares the stochastic simulation and the mean-field solution in this case. In contrast to the previous figures, the stochastic wave translates with lower speed than its deterministic counterpart. This apparent contradiction has a simple, intuitive explanation.

In general, given $\alpha > \beta$, we can expect the stochastic wave to be on the right of the corresponding deterministic solution for any values of $\lambda_1$ and $\lambda_2$. Since the population sizes in the deterministic solution are continuous variables, states that would be otherwise empty in the stochastic system always have a small non-zero density. Wave propagation is largely affected by these levels, particularly the left-most occupied ones, since pairwise copying enables large and non-local jumps. Consequently, `jumps back', i.e. to any earlier state, are always possible in the mean-field solution -- even if the equivalent stochastic jumps are not. Therefore, interactions that transport individuals to the left are more frequent in the deterministic than in the stochastic system. Numerical results for different values of $\lambda_1$ and $\lambda_2$ confirm this prediction (data not shown).

Besides the dissimilar speed of the stochastic wave, Fig.~\ref{fig:same_lambda} exhibits a flattening of the averaged curve as time progresses. This is caused by the diffusion of the wavefronts' position for different simulations at the same instant of time $t$. Despite the curves being similar in shape for each simulation, their position at a given time is different, resulting in an averaged curve that flattens as time progresses, indicating a diffusion in the co-moving frame of reference. In~\citet{BREUER1994259}, the authors investigate fluctuations on the stochastic wavefront of a system of particles $X$ that diffuse and react locally according to $X+Y \rightleftarrows 2X$ with constant concentration of particles $Y$. Their analysis is restricted to a continuous approximation of the lattice system, but notably, they observe the same flattening of the average curve as we. They suggest that this deviation arises from an asymmetrical influence of the large fluctuations of the wave front position upon its average drift, which are neglected by the linear noise approximation.

\begin{figure}[t]
\includegraphics[width=8.6cm]{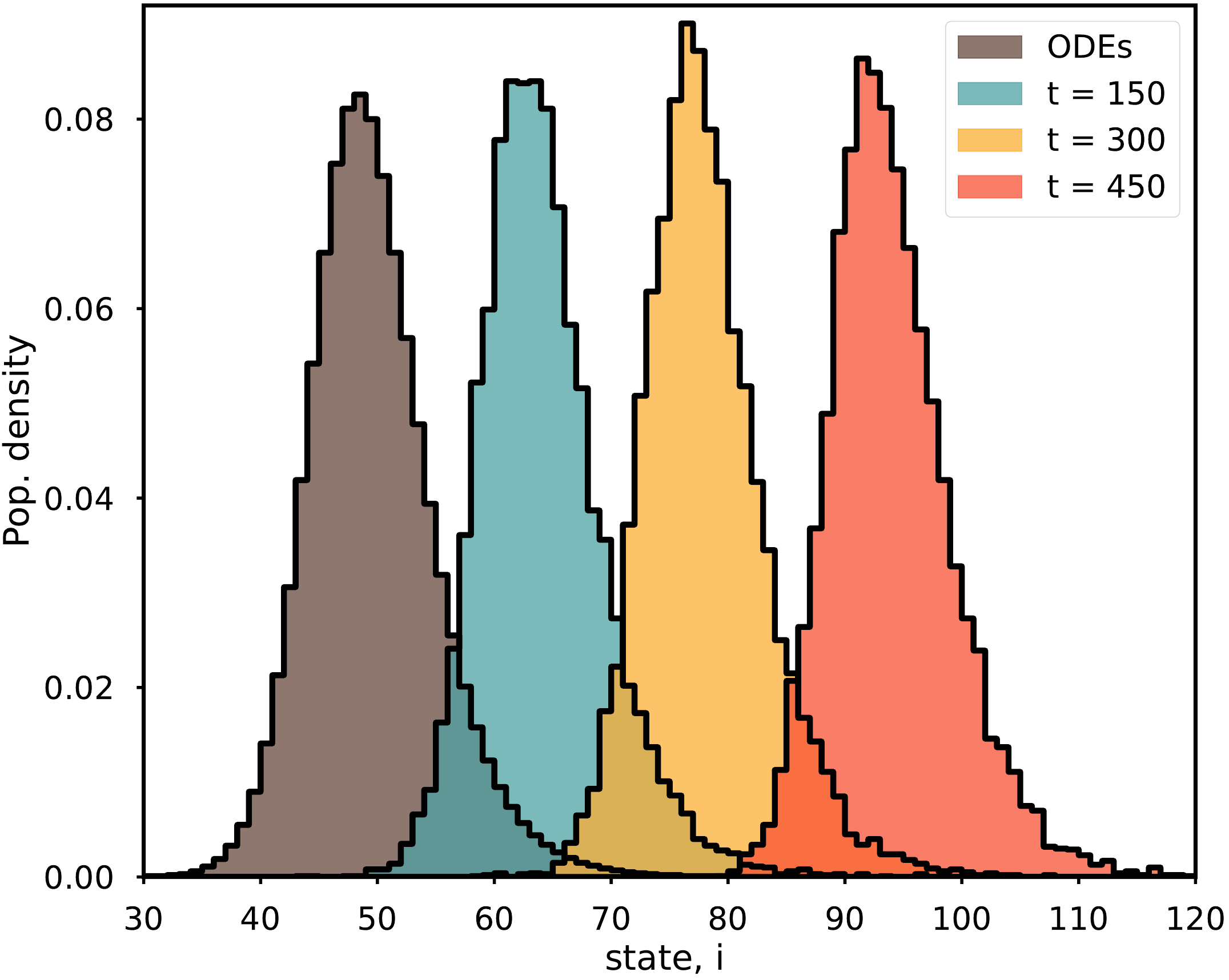}
\caption{\label{fig:stationary_wave} The stochastic wave (turquoise, yellow and orange) and the solution of Eq.~\eqref{eq:mf_ode_first} (brown) plotted at different times, $t$. The stochastic wave slowly drifts, whilst the deterministic solution remains static. Parameters: $N=10^4$, $\alpha = 1$, $\beta = 0$, $\lambda_1 = 4$, $\lambda_2 = 1$. Initial conditions as explained in text.}
\end{figure}


As we have seen, given our choice of interaction-function, $f$, and $\alpha > \beta$, the wave propagates to the right if the outward update and pairwise copying are the only interactions ($\lambda_2=0$), and to the left if individual updates to both directions occur with same rate ($\lambda_1=\lambda_2=\lambda$). One might then ask if there is a choice of parameters that balances right and left movement and keeps the wave static. Linearising the mean-field ODEs and setting $v_c=0$ gives the relation $\zeta = (\sqrt{\lambda_1} - \sqrt{\lambda_2})^2$ (see Appendix~\ref{apx:stationary_wave} for details). In Fig.~\ref{fig:stationary_wave} we show the solution of the deterministic system of equations and the stochastic simulation for this choice of parameters. To obtain the figure, we started with the initial condition $n_{50}(0)=N$ for the deterministic equations and then used the stationary distribution after the system had time to converge as initial condition for the stochastic system. We note a qualitative difference between the stationary mean-field solution and the stochastic simulation, which slowly drifts to higher states.

A key result so far has been the different speeds with which the stochastic and deterministic waves propagate. Numerical simulations for other choices of $f(z)$ satisfying the conditions $\lim_{z\to-\infty} f(z) = \alpha$ and $\lim_{z\to\infty}f(z) = \beta$ have shown similar behaviour (see Appendix~\ref{apx:general_f} for some examples). In the next section, we obtain a correction for the speed of the stochastic system and the diffusion rate of the wave's position between realisations of the system by making use of the cut-off theory  developed by~\citet{Brunet_1997}. These effects are shown to decay as $(\log N)^{-2}$ and $(\log N)^{-3}$ with the population size, $N$.

\subsection{The wavefront speed correction}
In this section, we analyse the model further with the aid of the cut-off method to obtain an analytical estimate of the speed of propagation of the stochastic wavefront. The cut-off method, developed by~\citet{Brunet_1997}, mimics the discrete nature of the population density in the stochastic wavefront by defining a cut-off on the deterministic equation in which the density at a given point is set to zero if it takes values below the cut-off threshold. Following a similar approach, we derive the speed correction for spatially-discrete variables and compare the results to numerical simulations.

In the previous section, we observed that the stochastic wavefronts are slower than the deterministic solution if $\lambda_1=\lambda_2=\lambda$, for our choice of $f(z)$ and $\alpha > \beta$. The opposite has been observed if $\lambda_2 = 0$, but $\lambda_1 \neq 0$. Despite this apparent dissonant behaviour, if we substitute the propagation speed of the stochastic wave into Eq.~\eqref{eq:speed_on_f}, we get a complex value of $\gamma$ in both cases.  In the deterministic model, solutions with complex $\gamma$ are not physically possible -- or meaningful -- as the densities become negative. 


For our choice of $f(z) = \alpha\theta(-z) + \beta\theta(z)$ and $\alpha > \beta$, the copying mechanism creates a bias towards the left, such that the position of the left-most occupied states dictate wave propagation, independently of its direction, rather than the bulk behind it. The position of the best informed individual dictates the position of the bulk, as copies of states at the front of the wave are more frequent than copies of positions at the back. Additionally, in the original information spread model~\cite{Funk2009} with $\lambda_2=0$, once the best-informed individual fades from $k$ to $k+1$, the $k$-quality information is lost and cannot be recovered.

This remarkable dependency on the left-most occupied sites of the wave allows us to focus our attention on the fluctuations in its far front and ignore the more complicated aspects of the highly-correlated noise brought about by non-local interactions. Combining the discreteness and pulled characteristic of the stochastic wave,~\citet{Brunet_1997} suggested the implementation of a cut-off in the corresponding PDE description of the system to compute a correction of the wavefront propagation speed. In this framework, the density solution is set to zero if it reaches a threshold value, $\varepsilon_N$. The adoption of a cut-off in the deterministic solution circumvents the issue of non-zero densities infinitely far in front of the wave, which causes the change in wave speed compared to the stochastic model. A natural choice is $\varepsilon_N = \frac{1}{N}$, to match the discreteness scale of the stochastic wave. 

We will follow their work in~\cite{Brunet_1997} to derive a similar wave speed correction for the discrete system of equations, Eq.~\eqref{eq:mf_ode}, for the variable $U_k(t)$, the cumulative sum of subpopulations in states $i \leq k$. A similar derivation is presented in~\cite{PhysRevE.66.036206} for the model studied in~\cite{BREUER1994259} which yields a discrete logistic KPP-equation, with $\lambda_1=\lambda_2=1$. Recall, we have assumed $f(z) = \alpha\theta(-z) + \beta\theta(z)$ and $\alpha > \beta$. The same derivation holds for $\alpha < \beta$ by interchanging $\lambda_1$ and $\lambda_2$, and reflecting the state-space such that left becomes right and vice-versa.

We first highlight the observed property that the population ``bulk'' maintains its shape as it translates through the states. This is a reasonable assumption if the rearranging of individuals within the bulk is much faster than the movement of the bulk itself. 

\begin{figure}[t]
\includegraphics[width=8.6cm]{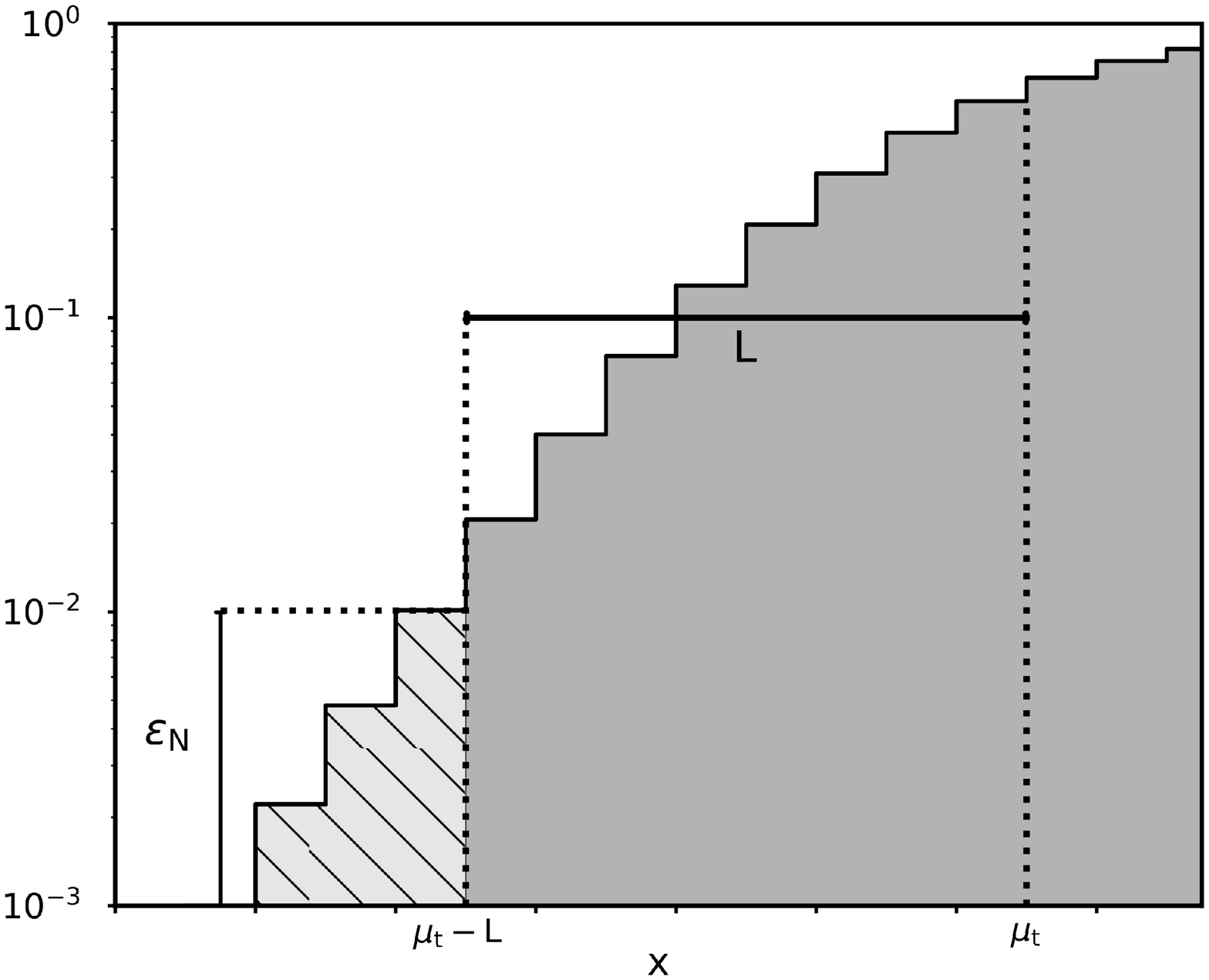}
\caption{\label{fig:cutoff_schematic} A schematic representation of the definitions adopted in the cut-off approach to estimate the speed of propagation of the stochastic wavefront. The deterministic solution is set to zero if it assumes values below the cut-off threshold, $\varepsilon_N$ (hatched region). This occurs at a distance $L$ from the reference position, $\mu_t$. Note the log-scale in the y-axis.}
\end{figure}

If $v$ is the signed speed of the wavefront, we may define the co-moving coordinate $x = k-vt$, where $k$ is our discrete variable that gives the state of an individual. Let $w_v(x)$ be the shape of the wavefront that propagates with speed $v$. Without loss of generality, we can define a reference position, $\mu_t$, such that $w_v(\mu_t) = \frac{1}{2}$. We can then define a cut-off distance $L > 0$ where $w_v(\mu_t - L) = \varepsilon_N$ and specify that $w_v(\mu_t-y) = 0$, if $y > L$. Fig.~\ref{fig:cutoff_schematic} gives a schematic picture of our definitions.

For simplicity, we may assume $k\in\mathbb{Z}$. We then look for travelling waves solutions $w_v(x)$ such that $U_k(t) = w_v(k - vt)$ is a solution to the cut-off problem,

\begin{widetext}
\begin{equation}\label{eq:cut-off}
    \begin{cases}
    \displaystyle-v\frac{dw_v}{d x} = \lambda_1 w_v(x-1) - (\lambda_1+\lambda_2)w_v(x) + \lambda_2 w_v(x+1)+ \zeta w_v(x) (1-w_v(x)),& x > \mu_t-L,\\
    \hspace{.25cm} w_v(x) = \varepsilon_N, & x = \mu_t - L,\\
    \hspace{.25cm} 0, &\text{otherwise}.
    \end{cases}
\end{equation}
\end{widetext}

As discussed, the asymptotic behaviour of the front for $x \to -\infty$, when the density is close to zero, determines the speed of propagation of the wavefront. Hence, we can look for solutions of the linearised equation. For $x > \mu_t - L$, the ansatz $w_v(x) = Ce^{\gamma x} = Ce^{\gamma(k-vt)}$ gives the following relationship between $\gamma$ and $v$, when substituted in Eq.~\eqref{eq:cut-off},
\begin{equation}\label{eq:speed}
    -\gamma v = \lambda_1 (e^{-\gamma}-1) + \lambda_2(e^\gamma-1) + \zeta.
\end{equation}

The values of the pairs $(v,\gamma)$ that satisfy the above expression can be obtained using the geometric method, i.e., by taking $v$ as a parameter and graphing the functions $g(\gamma) = -\gamma v$ and $h(\gamma) = \lambda_1 (e^{-\gamma}-1) + \lambda_2(e^\gamma-1) + \zeta$ to find the points where the two functions intersect. The left-hand-side of the expression defines a family of functions indexed by $v$, whereas the right-hand-side is independent of $v$. Thus, depending on the value of $v$, the above expression can have zero, one, or two real solutions for the pair $(v,\gamma)$. In particular, the point where the solution is unique defines the critical values $(v_c,\gamma_c)$. Note that the sign of $v$ and $v_c$ gives the direction of wave propagation, which depends on the system's parameters.

The deterministic wave will propagate with the critical speed $v_c$, whilst the stochastic wave propagates with speed $v$ in the parameter region where $\gamma$ assumes complex values. Let $\Delta = v_c - v$, i.e., the difference in speed between the deterministic and the stochastic wave and $\gamma = \gamma_r + i\gamma_i$, where $\gamma_r$ is the real part and $\gamma_i$ the imaginary part of the complex $\gamma$ corresponding to the speed $v$ of the stochastic wave. The solutions for the linearised equations then take the form,
\begin{equation}
    w_v(x) = [A_v\sin(\gamma_i x + \phi_v) + o(1)] e^{\gamma_r x}
\end{equation}
for $-L < x \ll 0$ and $v(\gamma) = v(\gamma_r + i\gamma_i)$ as given by Eq.~\eqref{eq:speed}, and constants $A_v$ and $\phi_v$. By definition, $v_c = v(\gamma_c)$ and $v'(\gamma_c) = 0$. If $\Delta$ is small, the speed is close to critical. Expanding $v(\gamma)$ up to second order around $\gamma_r$ yields
\begin{eqnarray}\label{eq:v_gamma}
    v(\gamma_r + i\gamma_i) &&= v(\gamma_r) + i\gamma_i v'(\gamma_r) - \frac{\gamma_i^2 }{2}v''(\gamma_r) + o(\gamma_i^3),\nonumber\\
    &&= v_c - \Delta\quad\text{(by definition)}.
\end{eqnarray}

Comparing the first and second lines of Eq.~\eqref{eq:v_gamma} gives
\begin{equation}
    \gamma_r = \gamma_c, \hspace{1cm} \gamma_i^2 = \frac{2\Delta}{v''(\gamma_c)}.
\end{equation}

We now use the translational invariance to set $\phi_v = 0$, and noticing that $\lim_{\Delta\to 0} A_{v}\sin(\gamma_i z) = o(1)$, since the sine factor is absent for waves with speed $v_c$ because $\gamma$ is real, we get
\begin{eqnarray}
    w_{v_c-\Delta}(x) &&= \frac{A}{\gamma_i}\sin\left(\gamma_i x\right)e^{\gamma_r x}\nonumber\\
    &&= A\sqrt{\frac{v''(\gamma_c)}{2\Delta}}\sin\left(\sqrt{\frac{2\Delta}{v''(\gamma_c)}}x\right)e^{\gamma_c x},
\end{eqnarray}
where $A$ is a constant that depends on initial conditions.

By assumption, this solution must be close to zero when $z \sim -L$, i.e.,
\begin{equation*}
    -\sqrt{\frac{2\Delta}{v''(\gamma_c)}}L \sim -\pi\Longrightarrow \Delta = \frac{\pi^2v''(\gamma_c)}{2L^2}. 
\end{equation*}

Furthermore, since the sine function varies much more slowly than the exponential, and $w_{v_c-\Delta} \sim \varepsilon_N$ for $z \sim -L$, we expect $e^{-\gamma_c L} \sim \varepsilon_N$, i.e., $L \sim -\frac{\log \varepsilon_N}{\gamma_c}$. Therefore,
\begin{equation}\label{eq:speed_correction}
    v = v_c - \Delta \sim v_c - \frac{\pi^2\gamma_c^2v''(\gamma_c)}{2\log^2\varepsilon_N} = v_c - \frac{\pi^2\gamma_c^2v''(\gamma_c)}{2\log^2 N},
\end{equation}
if we take $\varepsilon_N = N^{-1}$. 

Eq.~\eqref{eq:speed_correction} is the same expression obtained by~\citet{Brunet_1997}. The difference here from the usual KPP equation is the function $v(\gamma)$ (given by $v(\gamma) = \gamma\lambda + \frac{\zeta}{\gamma}$ in the PDE case) that gives the speed of propagation as a function of the exponent $\gamma$. \citet{Dumortier_2007} showed that the first-order correction in the propagation speed predicted by Brunet and Derrida holds for a large class of cut-off functions. 

The dependence of the correction, Eq.~\eqref{eq:speed_correction}, on the parameters is hidden inside $\gamma_c$ and $v''(\gamma_c)$. To obtain insight on how the parameters affect the speed correction, note that the deterministic approximation will be better if $|\gamma_c^2v''(\gamma_c)|$ is small. From Eq.~\eqref{eq:speed}, it follows that $\gamma_c^2v''(\gamma_c) = -\gamma_c(\lambda_1 e^{-\gamma_c} + \lambda_2 e^{\gamma_c})$. Hence, $|\Delta|\to 0$ as $\gamma_c\to 0$. In particular, $\gamma_c = 0 \iff \zeta = 0$, i.e., either particle interaction is symmetric or they do not interact. This is consistent with the result in Appendix~\ref{apx:symmetric}, where we showed that the deterministic approximation describes the average of the simulations exactly if particle interactions are symmetric. 

The dependence of $\gamma_c$ on the parameters is expressed through the following transcendental equation
\begin{equation}\label{eq:gamma_transcendental}
    \lambda_1e^{-\gamma_c}(1+\gamma_c) + \lambda_2e^{\gamma_c}(1-\gamma_c) + \zeta - \lambda_1-\lambda_2 = 0.
\end{equation}

To get some insight on how $\gamma_c$ changes with the parameters for $\gamma_c$ near zero, we Taylor expand Eq.~\ref{eq:gamma_transcendental} around $\gamma = 0$ up to second order to get
\begin{equation}
    \gamma_c^2 = \frac{2\zeta}{\lambda_1+\lambda_2}.
\end{equation}

Thus, in order to get a better match between the deterministic and stochastic solutions we need $\zeta \to 0$, so that the non-local interactions are absent or completely symmetrical. Alternatively, we could increase the values of $\lambda_1$ and/or $\lambda_2$. Since $\lambda_1$ and $\lambda_2$ are local, spontaneous interactions, they break the spatial correlation $\langle n_i(t)n_j(t)\rangle$ created by the non-local interactions, allowing the model to be better approximated by the mean-field solution.

\begin{figure}[t]
\includegraphics[width=8.6cm]{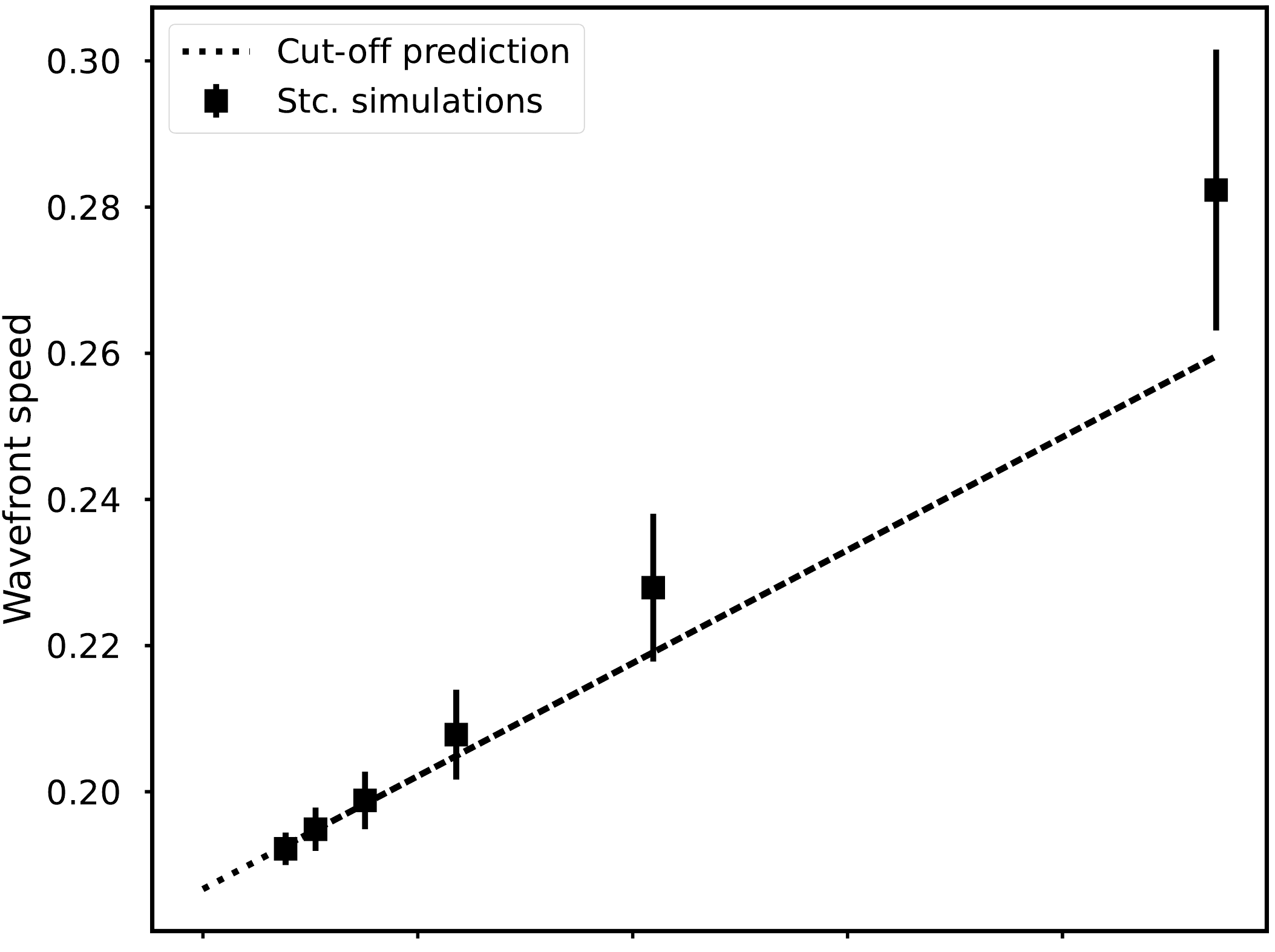}
\includegraphics[width=8.6cm]{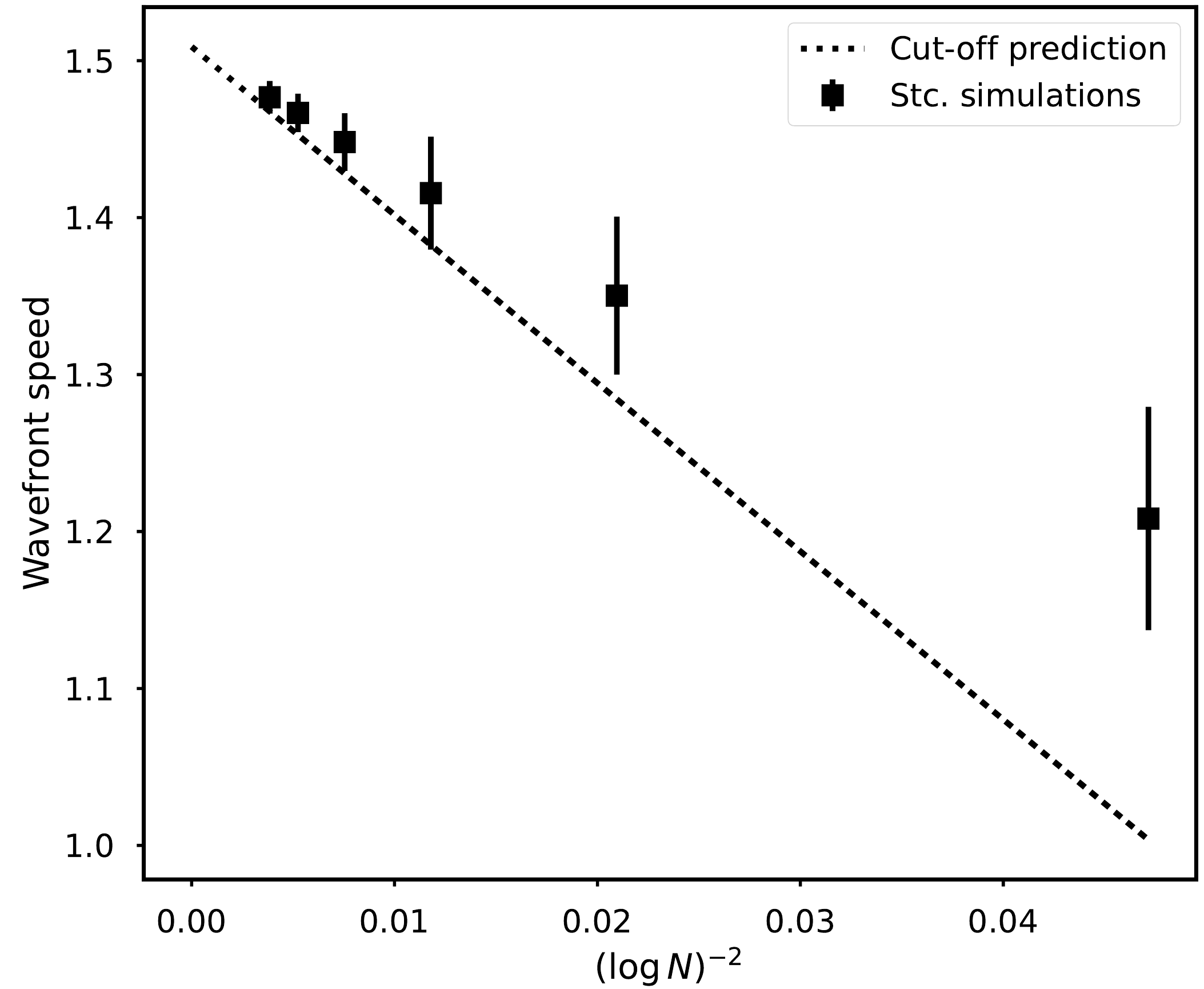}
\caption{\label{fig:speed_correction} The speed of propagation of the stochastic wavefront (squares) for different population sizes. The error bars give the standard deviation of the simulations for each population size. Upper figure: $\alpha = 0.5$, $\beta = 0$, $\lambda_1=1$, $\lambda_2=0$. Bottom figure: $\alpha = 1, \lambda_1 = \lambda_2 = \frac{1}{2}$, $\beta = 0$.}
\end{figure}

Note the anomalous $(\log N)^{-2}$ scaling of the speed correction with the population size, $N$. This scaling indicates a much slower decay of the stochastic effects of small population sizes than the usual $N^{-\frac{1}{2}}$ scaling. In Fig.~\ref{fig:speed_correction} we compare this prediction with the speed of propagation of the stochastic wavefront for different population sizes, $N$, for the cases $\lambda_2 = 0$ (the original model) and $\lambda_1=\lambda_2=\lambda$. Although the $(\log N)^{-2}$ scaling seems to be correct, the scaling constant (the slope of the line in the plot) seems to disagree with the one predicted by the cut-off approach, suggesting the constant in Eq.~\eqref{eq:speed_correction} may be different. Alternatively, the limitations of the cut-off method for small $N$ and the slow scaling of the correction suggest that the agreement between numerical results and the theoretical prediction may improve for larger population sizes.

Besides the difference in wave speeds, we also observed in Sec.~\ref{sec:numerical_results} the fluctuation of the wavefront's position as a result of stochastic effects. Let $D_N$ be the diffusion of the wavefront position from realisation to realisation,
\begin{eqnarray}
    D_N = \lim_{t\to\infty}\frac{\langle \mu_t^2 \rangle - \langle \mu_t\rangle^2}{t},
\end{eqnarray}
where $\langle\hspace{.05cm} \boldsymbol{\cdot}\hspace{.05cm} \rangle$ is the average over realisations.

\begin{figure}[t]
\includegraphics[width=8.6cm]{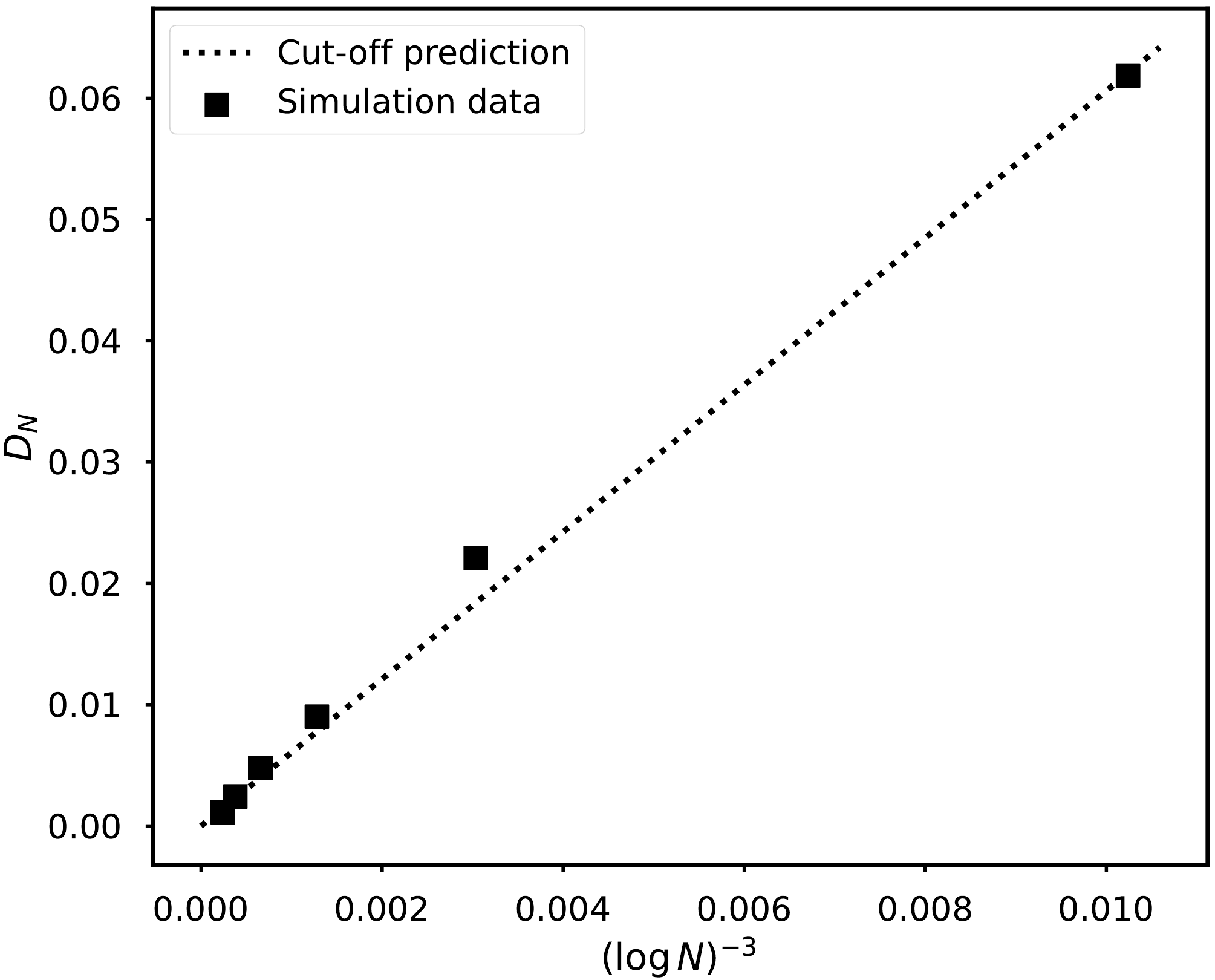}
\caption{\label{fig:dif_correction} The diffusion constant of the stochastic wavefront (squares) for different population sizes. Parameters: $\alpha = 0.5$, $\beta = 0$, $\lambda_1 = 1$ $\lambda_2 = 0$.}
\end{figure}

A result derived by~\citet{Brunet_1997} for KPP-equations with cut-off also suggest that the diffusion rate, $D_N$, scales with the population size as 
\begin{eqnarray}
    D_N = \frac{\pi^4 \gamma_c v''(\gamma_c)}{3\log^3 N},
\end{eqnarray}
where $\gamma_c$ and $v''(\gamma)$ are model dependent. Note again the extremely slow scaling of this diffusion rate (of order $(\log N)^{-3}$) with the population size $N$. In Fig.~\ref{fig:dif_correction} we verify this prediction by plotting the diffusion constant obtained from the stochastic simulations for different population sizes, $N$. As before, the $(\log N)^{-3}$ decay seems to be correct, as well as the constant of proportionality.

The anomalous scalings observed indicate that fluctuations in the wavefront's position and differences in speed of propagation with respect to predictions of the mean-field approximation are observed even for large population sizes. This effect of stochasticity can have critical real-world consequences. Particularly, in the case $\lambda_2 = 0$, corresponding to the Muller's ratchet and the original awareness spread model, the stochastic wave propagates faster than its deterministic counterpart even for very large population sizes. This suggests that the accumulation of deleterious mutations in a population reduces its overall fitness and makes it more susceptible to extinction, whilst the loss of awareness reduces the adoption of non-pharmaceutical interventions in epidemics, leading to increased disease transmission. Moreover, these results affects predictions on the success of invading species, which frequently model interactions between species by deterministic KPP-equations~\cite{Holmes_1994}, by altering the speed with which they occupy a new environment.

Despite the good agreement of Brunet and Derrida's corrections observed for many models, the apparent disagreement with the constant observed for our model is in line with previous research~\cite{Brunet_2006,doi:10.1063/1.479923,Kessler1998}. One reason could be the contributions from all space scales to the fluctuations, not just from the very front or the very end of the wave, in contrast to most branching particle systems studied by Brunet and Derrida. These contributions lead to stronger spatial-scale correlations in our model, which are expected to lead to larger disagreements with the moment-closure approximation. 
A different scaling factor is also observed by~\citet{Brunet_2006} in a model of branching particles with selection and stronger spatial correlations.

In the next section, we discuss and give insight on other classes of models that we expect to show similar disagreement between the
stochastic and the mean-field solutions.

\section{Discussion}\label{sec:discussion}

Mean-field approximations are cornerstones of mathematical biology. They are often used to derive simplified deterministic equations for stochastic systems. These approximations usually describe well systems with many particles and the averages of numerous simulations. Despite this, stochasticity is often the origin of fascinating effects not predicted by the mean-field equations.

In this paper, we investigated a generalised model for awareness spread in a population. This generalisation can be used as a toy model in several areas of research, ranging from ecology to epidemics, cell motility and evolution. We showed that this model, under a useful change of variables, is described by a KPP-equation and exhibit wavefront solutions. We observed that the deterministic solution obtained through the mean-field approximation of this model does not agree well with the average of numerous stochastic simulations. The disagreement comes from finite-size effects that cause fluctuations in the position of the stochastic wave and change its speed of propagation. These stochastic effects are shown, through the application of the cut-off theory and numerical simulations, to decay with the population size $N$ as $(\log N)^{-3}$ and $(\log N)^{-2}$, respectively, much slower than the frequently observed $N^{-\frac{1}{2}}$ scaling. Hence, the accumulation of deleterious mutations in a Muller's ratchet, leading to a faster fitness loss and increased risk of extinction, and the loss of awareness in populations are effects observed to occur faster in the stochastic model than predicted by the deterministic theory, even for larger populations. Although our results were obtained for the choice of $f(z) = \alpha \cdot \theta(-z) + \beta \cdot \theta(z)$, numerical simulations for other choices of $f(z)$ such that $\lim_{z\to-\infty} f(z) = \alpha$ and $\lim_{z\to\infty}f(z) = \beta$ have been consistent with our observations. 

The versatility of the cut-off theory has led to a considerable area of research on how a cut-off affects the speed of propagation of deterministic reaction-diffusion equations of the KPP type~\cite{PANJA200487,PhysRevLett.87.238303,PhysRevE.65.057202,PhysRevE.68.065202}, to cite but a few. Some work has been done to apply this methodology to on-lattice systems and some specific interacting-particle stochastic models with local interactions~\cite{PhysRevE.66.036206,Kessler1998}. Yet, the question of what general class of individual-based stochastic models are expected to present this anomalous scaling is still unanswered. Here, we tried to provide some insights on this problem. We hypothesize that any model with wave propagation lead by particles at its front and linear growth for small densities should have the anomalous $\log^2N$ scaling observed for our model, although models with stronger, non-local, spatial interactions might have a different scaling constant than predicted by the cut-off theory. Examples explored in the literature are Branching Brownian motion and interacting-particle systems on lattices. These are widely used in chemistry, physics and mathematical biology modelling -- in areas that range from ecology to cell motility, epidemiology, evolution and collective behaviour. The ubiquity of these models in research highlights the importance of studying and understanding the effects of stochasticity in wave propagation to pinpoint the extension to which results derived from established deterministic models are valid. 

Further work would be welcomed on extending the classes of models that show anomalous scaling with the population size. Another direction for research would be to explain why some models agree well with the speed correction derived by Brunet and Derrida, whilst others present a different scaling constant. This has proven to be difficult. Tentative research has been developed in this aspect~\cite{PANJA200487} with a probabilistic investigation of the foremost occupied site. The theory is, however, not predictive, reducing the extent of its applicability. Finally, the derivation of more and improved analytical results for spatially discrete stochastic systems based on the cut-off framework would be a fruitful direction of research, as most of the analysis developed so far has concerned PDEs and continuous space.

\begin{acknowledgments}
\noindent A.S. is supported by a scholarship from the EPSRC Centre for Doctoral Training in Statistical Applied Mathematics at Bath (SAMBa), under the project EP/S022945/1.
\end{acknowledgments}

\appendix

\section{Symmetric jumps}\label{apx:symmetric}
In this appendix we show that, when pairwise copying is unbiased, i.e. $f(x) = f(-x)$, then the deterministic approximation describes the average of the simulations exactly.

Recall that Eq.~\eqref{eq:master_equation} gives the master equation of our general model. Multiplying this equation by $n_k$ and summing over the state space gives
\begin{eqnarray}
    \frac{d\langle n_k\rangle}{dt} =&& \sum_{i} \left\langle n_k n_i \frac{f(k-i)}{N} \right\rangle - \sum_{i} \left\langle n_k n_i \frac{f(i-k)}{N} \right\rangle\nonumber\\
    &&+ \lambda_1 \langle n_{k-1}\rangle-(\lambda_1+\lambda_2) \langle n_k\rangle + \lambda_2 \langle n_{k+1}\rangle.
\end{eqnarray}

By assumption, $f(k-i) = f(i-k)$. Hence, the two sums cancel, and we are left with
\begin{equation}
    \frac{d\langle n_k\rangle}{dt} = \lambda_1\langle n_{k-1}\rangle - (\lambda_1+\lambda_2) \langle n_k\rangle + \lambda_2 \langle n_{k+1}\rangle.
\end{equation}

As the equations depend solely on first-order moments, no moment closure approximation is required and the result is exact. The average over many simulations and the solution of this system of equations must agree well. Therefore, asymmetric jump rates are a necessary condition to observe disagreement.

\section{Linear functions of the distance}\label{apx:polynomial}
In this appendix, we investigate the case where $f(z)$ increases linearly with the difference between states and show that the system achieves a stationary distribution rather than wave propagation.

Let us analyse a modification of the original model in~\cite{Funk2009} where two-body jumps occur only to better qualities of information with rate that depends linearly on the distance between two levels. This case is described by $f(z) = \alpha\cdot\theta(-z)$, where $\theta$ is the Heaviside function, and $\lambda_2 = 0$. For this choice of $f(z)$, the discrete system of mean-field equations becomes
\begin{equation*}
    \begin{cases}
            \displaystyle\frac{\text{d} n_0}{\text{d}t} = -\lambda_1 n_0 + \frac{\alpha}{N}n_0\sum_{j=0}^\infty j\cdot n_j,& k = 0,\\
            \displaystyle\frac{\text{d}n_k}{\text{d}t} = -\lambda_1 n_k + \lambda_1 n_{k-1} + \frac{\alpha}{N}n_k\sum_{j=0}^\infty n_j(j-k),& k > 0,
    \end{cases}
\end{equation*}
where we have used the mean-field approximation $\langle n_in_j\rangle = \langle n_i\rangle \langle n_j \rangle$, and written $\langle n_i \rangle \mapsto n_i$. Denoting $\overline{k} = \frac{1}{N}\sum j\cdot n_j$, the average state in the population, the previous equations can be written as
\begin{equation*}
    \begin{cases}
            \displaystyle\frac{\text{d} n_0}{\text{d}t} = -\lambda_1 n_0 + \alpha n_0\overline{k},& k = 0,\\
            \displaystyle\frac{\text{d}n_k}{\text{d}t} = -\lambda_1 n_k + \lambda_1 n_{k-1} + \alpha n_k(\overline{k} - k),& k > 0.
    \end{cases}
\end{equation*}

To get the steady states, we set the derivatives to zero. The equation for $k = 0$ gives $n_0^{st} = 0$ or $\overline{k} = \frac{\lambda_1}{\alpha}$. The first condition gives $n_k^{st} = 0$, whereas the second gives
\begin{equation}
    n_k^{st} = \frac{\lambda_1}{\alpha k}n_{k-1}^{st}\Longrightarrow n_k^{st} = \left(\frac{\lambda_1}{\alpha}\right)^{k-k_0}\cdot\frac{1}{(k-k_0)!}n_{k_0}^{st},
\end{equation}
where $k_0$ corresponds to the first non-zero level in the initial condition of the system. In particular, $k_0 = 0$ for the Kronecker delta initial condition, $n_k = N\cdot \delta_{0,k}$. It follows that the distribution at the steady state is Poisson, with rate $\lambda_1/\alpha$ and support on $[k_0,\infty)$ . The normalisation condition then gives $n_{k_0}^{st} = Ne^{-\lambda_1/\alpha}$. Thus, taking $k_0 = 0$ for simplicity,
\begin{equation}
    n_k^{st} = N\cdot \left(\frac{\lambda_1}{\alpha}\right)^k\frac{e^{-\lambda_1/\alpha}}{k!} = N\cdot\text{Poiss}(\lambda_1/\alpha).
\end{equation}

Using linear analysis of the steady states, we can show that the trivial state is always unstable. The numerical solution of the discrete system of equations agrees well with this result. We note, however, that in the stochastic system, $\mathbb{P}[n_k(t) = 0] = 1$ as $t\to\infty$ for all $k$.

An interesting corollary of this Poisson distribution of the steady states is that the average level and variance of the population can be easily obtained to be $\overline{k} = \text{Var}(k) =  \frac{\lambda_1}{\alpha}$. A similar approach applied to any polynomial function $f(z)$ will render $\bar{f}^{st} = \frac{1}{N}\sum f(j)\cdot n_j^{st} = \frac{\lambda_1}{\alpha}$ with unstable trivial steady state since $\lim_{j\to \infty} f(j) = \infty$. Numerical results indicated good agreement with these predictions (not shown).

\section{Stationary deterministic wave}\label{apx:stationary_wave}
In this appendix, we find a relationship between the model's parameters that result in a stationary wave solution for Eq.~\eqref{eq:mf_ode}.

Assuming $\zeta = \alpha - \beta > 0$, i.e., the pairwise copying is biased towards left-movement, and that particles are allowed to move right with rate $\lambda_1$ and to move left with rate $\lambda_2$, we would expect $\lambda_1 > \lambda_2$ in the stationary system, if it exists, to counter-balance the $\lambda_2$ and $\zeta$ rates of movement to the left. But what is the exact relationship between the parameters?

In this case, the mean-field equations take the form
\begin{equation*}
    \frac{\text{d}U_k}{\text{d}t} = -(\lambda_1+\lambda_2)U_k + \lambda_1 U_{k-1} + \lambda_2 U_{k+1} + \zeta U_k (1 - U_k).
\end{equation*}

Looking for solutions of the form $U_k(t) = Ce^{\gamma(k-vt)}$, of the linearised equation, we get the following expression for the speed of propagation,
\begin{equation}
    -\gamma v = \zeta - (\lambda_1+\lambda_2) + \lambda_1 e^{-\gamma} + \lambda_2 e^{\gamma}.
\end{equation}

Note that, taking $\lambda_2 = 0$ gives the expression obtained for the model with only fading, and taking $\lambda_2 = \lambda_1$ gives the expression for the fading-and-finding model, as expected. This expression gives a family of solutions for $(\gamma,v)$ where the right and left sides of the equation intersect (geometric method). To find the critical speed, we differentiate both sides w.r.t $\gamma$ to obtain
\begin{equation}
    -v = -\lambda_1 e^{-\gamma} + \lambda_2 e^\gamma = 0\Longrightarrow e^\gamma = \sqrt{\frac{\lambda_1}{\lambda_2}}.
\end{equation}

Substituting back gives
\begin{align}
    0 &= \zeta - \lambda_1 - \lambda_2 + \lambda_1 \sqrt{\frac{\lambda_2}{\lambda_1}} + \lambda_2 \sqrt{\frac{\lambda_1}{\lambda_2}},\nonumber\\
    \zeta &= (\lambda_1 + \lambda_2 - 2\sqrt{\lambda_1\lambda_2})= (\sqrt{\lambda_1} - \sqrt{\lambda_2})^2.
\end{align}

\section{Stochastic waves for other choices of interaction-function $f(z)$}\label{apx:general_f}
In this appendix, we give examples of other choices of $f(z)$ satisfying the conditions $\lim_{z\to-\infty} f(z) = \alpha$ and $\lim_{z\to\infty}f(z) = \beta$ and show numerically that they exhibit similar behaviour as our choice $f(z) = \alpha\theta(-z) + \beta(z)$. For each figure, the corresponding function $f(z)$ is plotted as an inset.

\begin{figure}[h]
\includegraphics[width=.44\textwidth]{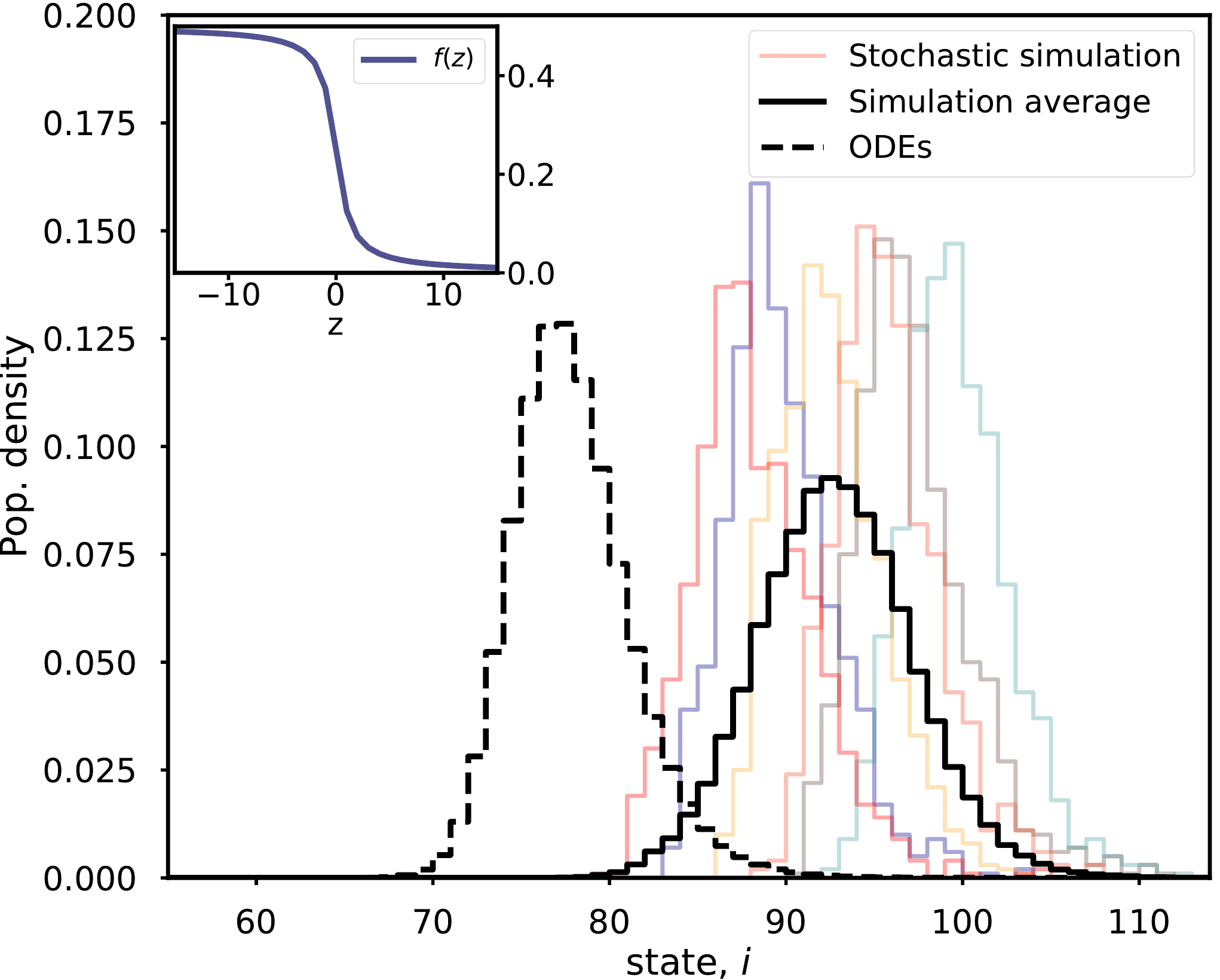}
\caption{\label{fig:apd1} Comparison between the deterministic solution of Eq.~\eqref{eq:moment_closure} (dashed) and stochastic simulations (colors) for $f(z) = \frac{1}{2\pi}\arctan(-z) + \frac{1}{4}$ (inset). For this $f(z)$, $\lim_{z\to-\infty} f(z) = \frac{1}{2}$ and $\lim_{z\to\infty}f(z) = 0$. Note the similarity with Fig.~\ref{fig:example_sim}. The parameter values used to generate the figure are: $N=10^3$, $t=400$, $\lambda_1 = 1$, $\lambda_2 = 0$. Initial condition: $n_{50}(0) = N$.}
\end{figure}

Fig.~\ref{fig:apd1} compares the stochastic simulations and the solution of Eq.~\eqref{eq:moment_closure} for $f(z) = \frac{1}{2\pi}\arctan(-z) + \frac{1}{4}$. For this $f(z)$, $f^{-} - f^{+} = \frac{1}{2}$ as in Fig.~\ref{fig:example_sim}. As expected, the stochastic waves are to the right of the deterministic solution and the waves' position from realisation to realisation fluctuates, flattening the average curve.

\begin{figure}[ht]
\includegraphics[width=.44\textwidth]{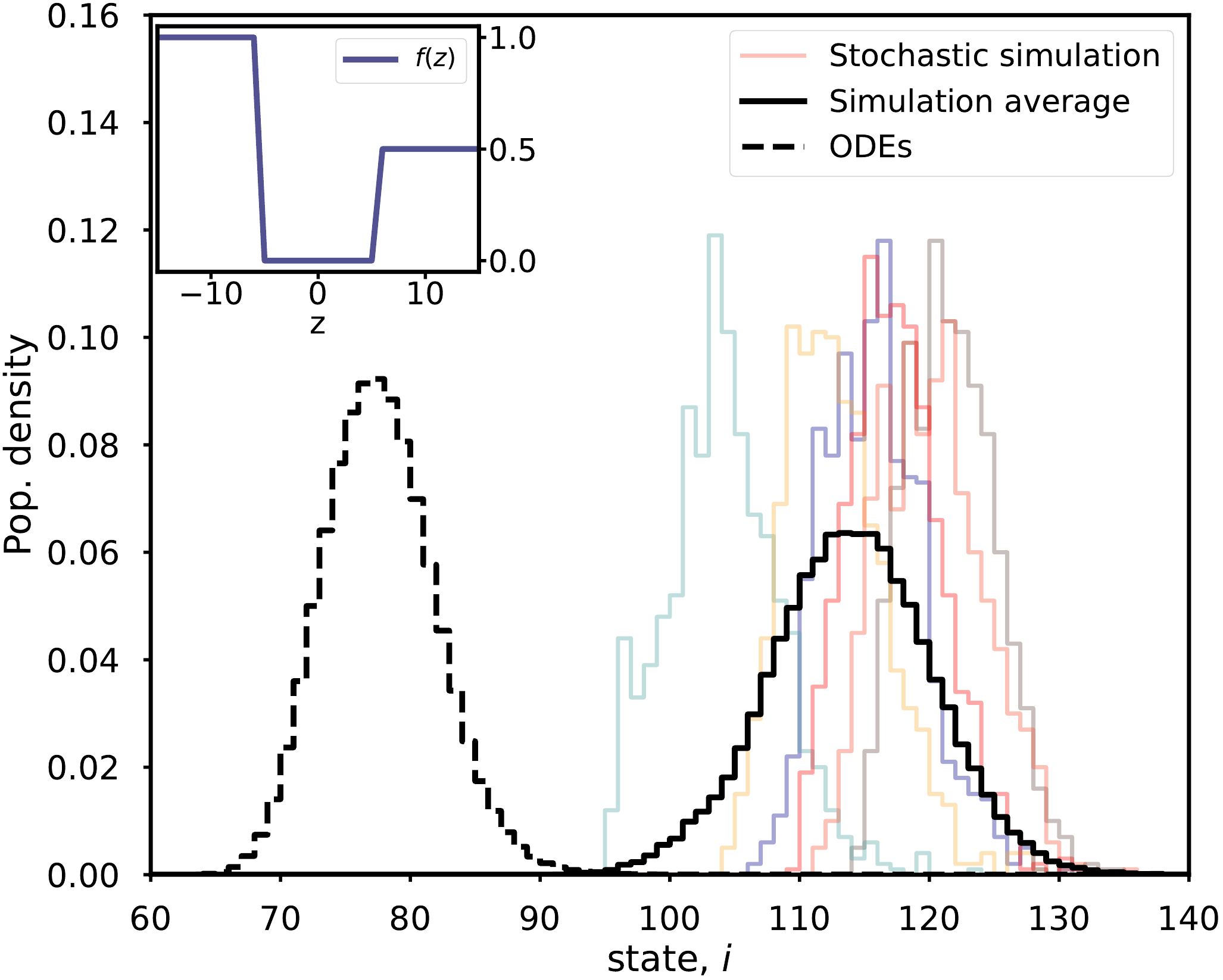}
\caption{\label{fig:apd2} Comparison between the deterministic solution of Eq.~\eqref{eq:moment_closure} (dashed) and stochastic simulations (colors) for $f(z) = \theta(-z-5) + \frac{1}{2}\theta(z-5)$ (inset). For this $f(z)$, $\lim_{z\to-\infty} f(z) = 1$ and $\lim_{z\to\infty}f(z) = \frac{1}{2}$, however, $f^{-}-f^{+} = \frac{1}{2}$ as before. The parameter values are: $N=10^3$, $t=400$, $\lambda_1 = 1$, $\lambda_2 = 0$. Initial condition: $n_{50}(0) = N$  (same as Fig.~\ref{fig:example_sim}).}
\end{figure}

In Fig.~\ref{fig:apd2} we compare the stochastic simulations and the solution of  Eq.~\eqref{eq:moment_closure} for $f(z) = \theta(-z-5) + \frac{1}{2}\theta(z-5)$. In this case, $f(z)$, $\lim_{z\to-\infty} f(z) = 1$ and $\lim_{z\to\infty}f(z) = \frac{1}{2}$, however, $f^{-}-f^{+} = \frac{1}{2}$ as before.


\begin{figure}[ht]
\includegraphics[width=.44\textwidth]{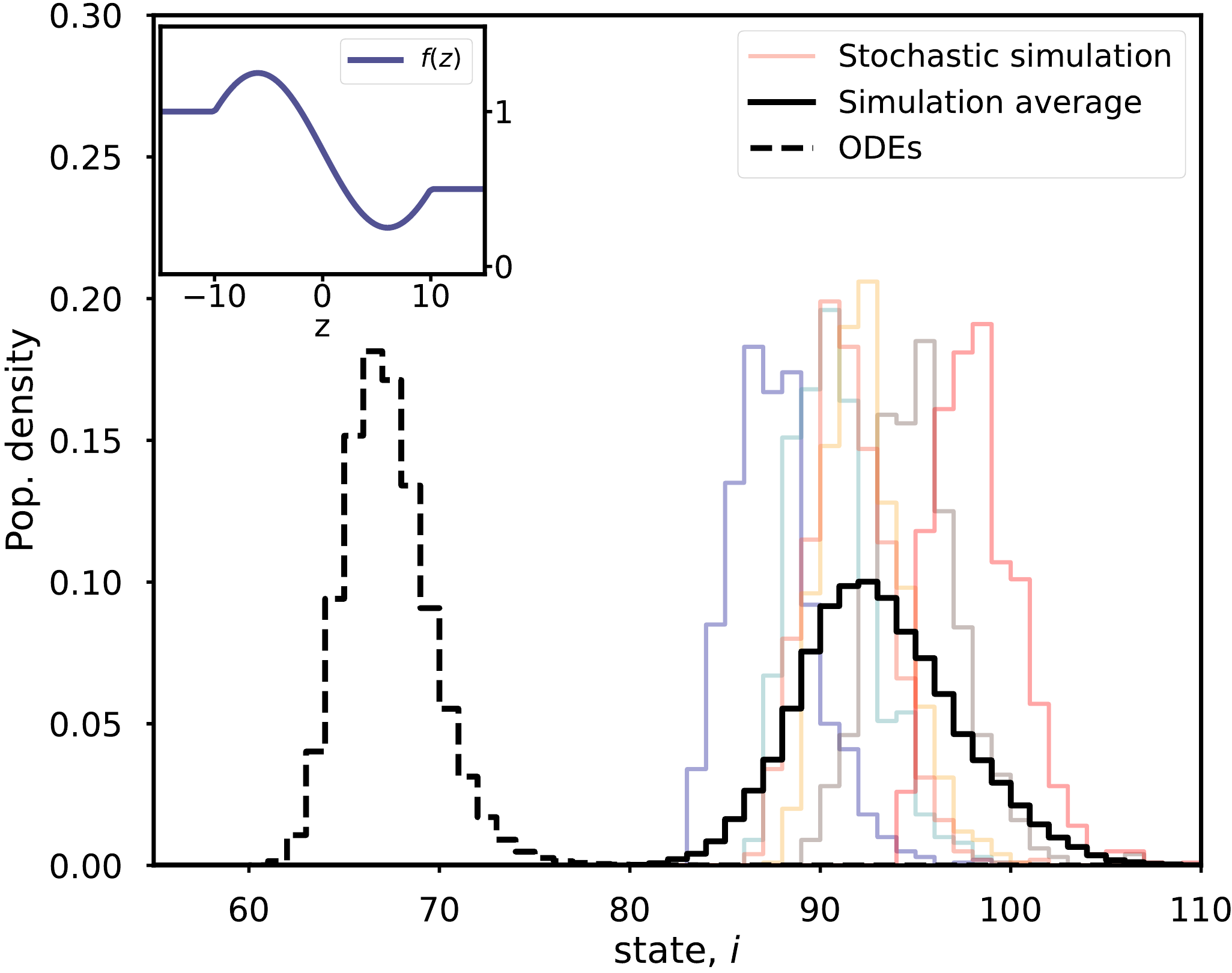}
\caption{\label{fig:apd3} Comparison between the deterministic solution of Eq.~\eqref{eq:moment_closure} (dashed) and stochastic simulations (colors) for the function $f(z)$ given by Eq.~\eqref{eq:f_apd3} (inset). For this $f(z)$, $\lim_{z\to-\infty} f(z) = 1$ and $\lim_{z\to\infty}f(z) = \frac{1}{2}$, however, $f^{-}-f^{+} =\frac{1}{2}$ as before. The parameter values are: $N=10^3$, $t=400$, $\lambda_1 = 1$, $\lambda_2 = 0$. Initial condition: $n_{50}(0) = N$.}
\end{figure}

Finally, in Fig.~\ref{fig:apd3} we make the same comparison, but this time considering the function,
\begin{eqnarray}\label{eq:f_apd3}
f(z) = \begin{cases}
       1, &\text{if}\ z < -10,\\
       \frac{1}{2}\sin(-\frac{\pi z}{12})+\frac{3}{4}, &\text{if}\ -10\leq z \leq 10,\\
       \frac{1}{2}, &\text{if}\ z > 10.
\end{cases}
\end{eqnarray}

As before, $\lim_{z\to-\infty} f(z) = 1$ and $\lim_{z\to\infty}f(z) = \frac{1}{2}$, hence, $f^{-}-f^{+} = \frac{1}{2}$. 

Fig.~\ref{fig:apd1},~\ref{fig:apd2} and~\ref{fig:apd3} hold several qualitative similarities to Fig.~\ref{fig:example_sim}. Particularly, the mismatching speed of propagation between the stochastic and deterministic models, the flattening of the average curve, and the fact that the stochastic wave is always to the right of the deterministic solution. These suggest that our modelling choice of interaction-function, $f(z)$, is appropriate and that our results can be extended for more general functions.


\bibliography{apssamp}

\providecommand{\noopsort}[1]{}\providecommand{\singleletter}[1]{#1}%
\begin{thebibliography}{46}%
\makeatletter
\providecommand \@ifxundefined [1]{%
 \@ifx{#1\undefined}
}%
\providecommand \@ifnum [1]{%
 \ifnum #1\expandafter \@firstoftwo
 \else \expandafter \@secondoftwo
 \fi
}%
\providecommand \@ifx [1]{%
 \ifx #1\expandafter \@firstoftwo
 \else \expandafter \@secondoftwo
 \fi
}%
\providecommand \natexlab [1]{#1}%
\providecommand \enquote  [1]{``#1''}%
\providecommand \bibnamefont  [1]{#1}%
\providecommand \bibfnamefont [1]{#1}%
\providecommand \citenamefont [1]{#1}%
\providecommand \href@noop [0]{\@secondoftwo}%
\providecommand \href [0]{\begingroup \@sanitize@url \@href}%
\providecommand \@href[1]{\@@startlink{#1}\@@href}%
\providecommand \@@href[1]{\endgroup#1\@@endlink}%
\providecommand \@sanitize@url [0]{\catcode `\\12\catcode `\$12\catcode
  `\&12\catcode `\#12\catcode `\^12\catcode `\_12\catcode `\%12\relax}%
\providecommand \@@startlink[1]{}%
\providecommand \@@endlink[0]{}%
\providecommand \url  [0]{\begingroup\@sanitize@url \@url }%
\providecommand \@url [1]{\endgroup\@href {#1}{\urlprefix }}%
\providecommand \urlprefix  [0]{URL }%
\providecommand \Eprint [0]{\href }%
\providecommand \doibase [0]{https://doi.org/}%
\providecommand \selectlanguage [0]{\@gobble}%
\providecommand \bibinfo  [0]{\@secondoftwo}%
\providecommand \bibfield  [0]{\@secondoftwo}%
\providecommand \translation [1]{[#1]}%
\providecommand \BibitemOpen [0]{}%
\providecommand \bibitemStop [0]{}%
\providecommand \bibitemNoStop [0]{.\EOS\space}%
\providecommand \EOS [0]{\spacefactor3000\relax}%
\providecommand \BibitemShut  [1]{\csname bibitem#1\endcsname}%
\let\auto@bib@innerbib\@empty
\bibitem [{\citenamefont {Lee}\ and\ \citenamefont
  {Yang}(1952)}]{PhysRev.87.410}%
  \BibitemOpen
  \bibfield  {author} {\bibinfo {author} {\bibfnamefont {T.~D.}\ \bibnamefont
  {Lee}}\ and\ \bibinfo {author} {\bibfnamefont {C.~N.}\ \bibnamefont {Yang}},\
  }\bibfield  {title} {\bibinfo {title} {Statistical theory of equations of
  state and phase transitions. ii. lattice gas and ising model},\ }\href
  {https://link.aps.org/doi/10.1103/PhysRev.87.410} {\bibfield  {journal}
  {\bibinfo  {journal} {Phys. Rev.}\ }\textbf {\bibinfo {volume} {87}},\
  \bibinfo {pages} {410} (\bibinfo {year} {1952})}\BibitemShut {NoStop}%
\bibitem [{\citenamefont {Holley}\ and\ \citenamefont
  {Liggett}(1975)}]{Holley_1975}%
  \BibitemOpen
  \bibfield  {author} {\bibinfo {author} {\bibfnamefont {R.~A.}\ \bibnamefont
  {Holley}}\ and\ \bibinfo {author} {\bibfnamefont {T.~M.}\ \bibnamefont
  {Liggett}},\ }\bibfield  {title} {\bibinfo {title} {Ergodic theorems for
  weakly interacting infinite systems and the voter model},\ }\href
  {https://doi.org/10.1214%2Faop%2F1176996306} {\bibfield  {journal} {\bibinfo
  {journal} {The Annals of Probability}\ }\textbf {\bibinfo {volume} {3}}
  (\bibinfo {year} {1975})}\BibitemShut {NoStop}%
\bibitem [{\citenamefont {Liggett}(1999)}]{Liggett1999}%
  \BibitemOpen
  \bibfield  {author} {\bibinfo {author} {\bibfnamefont {T.~M.}\ \bibnamefont
  {Liggett}},\ }\bibinfo {title} {Voter models},\ in\ \href
  {https://doi.org/10.1007/978-3-662-03990-8_3} {\emph {\bibinfo {booktitle}
  {Stochastic Interacting Systems: Contact, Voter and Exclusion Processes}}}\
  (\bibinfo  {publisher} {Springer Berlin Heidelberg},\ \bibinfo {address}
  {Berlin, Heidelberg},\ \bibinfo {year} {1999})\ pp.\ \bibinfo {pages}
  {139--208}\BibitemShut {NoStop}%
\bibitem [{\citenamefont {Rhodes}\ and\ \citenamefont
  {Anderson}(1996)}]{RHODES1996125}%
  \BibitemOpen
  \bibfield  {author} {\bibinfo {author} {\bibfnamefont {C.}~\bibnamefont
  {Rhodes}}\ and\ \bibinfo {author} {\bibfnamefont {R.}~\bibnamefont
  {Anderson}},\ }\bibfield  {title} {\bibinfo {title} {Persistence and dynamics
  in lattice models of epidemic spread},\ }\href
  {https://www.sciencedirect.com/science/article/pii/S0022519396900880}
  {\bibfield  {journal} {\bibinfo  {journal} {Journal of Theoretical Biology}\
  }\textbf {\bibinfo {volume} {180}},\ \bibinfo {pages} {125} (\bibinfo {year}
  {1996})}\BibitemShut {NoStop}%
\bibitem [{\citenamefont {Rhodes}\ and\ \citenamefont
  {Anderson}(1997)}]{RHODES1997101}%
  \BibitemOpen
  \bibfield  {author} {\bibinfo {author} {\bibfnamefont {C.}~\bibnamefont
  {Rhodes}}\ and\ \bibinfo {author} {\bibfnamefont {R.}~\bibnamefont
  {Anderson}},\ }\bibfield  {title} {\bibinfo {title} {Epidemic thresholds and
  vaccination in a lattice model of disease spread},\ }\href
  {https://www.sciencedirect.com/science/article/pii/S004058099791323X}
  {\bibfield  {journal} {\bibinfo  {journal} {Theoretical Population Biology}\
  }\textbf {\bibinfo {volume} {52}},\ \bibinfo {pages} {101} (\bibinfo {year}
  {1997})}\BibitemShut {NoStop}%
\bibitem [{\citenamefont {Grassberger}(1983)}]{GRASSBERGER1983157}%
  \BibitemOpen
  \bibfield  {author} {\bibinfo {author} {\bibfnamefont {P.}~\bibnamefont
  {Grassberger}},\ }\bibfield  {title} {\bibinfo {title} {On the critical
  behavior of the general epidemic process and dynamical percolation},\ }\href
  {https://www.sciencedirect.com/science/article/pii/0025556482900360}
  {\bibfield  {journal} {\bibinfo  {journal} {Mathematical Biosciences}\
  }\textbf {\bibinfo {volume} {63}},\ \bibinfo {pages} {157} (\bibinfo {year}
  {1983})}\BibitemShut {NoStop}%
\bibitem [{\citenamefont {Codling}\ \emph {et~al.}(2008)\citenamefont
  {Codling}, \citenamefont {Plank},\ and\ \citenamefont
  {Benhamou}}]{Codling_2008}%
  \BibitemOpen
  \bibfield  {author} {\bibinfo {author} {\bibfnamefont {E.~A.}\ \bibnamefont
  {Codling}}, \bibinfo {author} {\bibfnamefont {M.~J.}\ \bibnamefont {Plank}},\
  and\ \bibinfo {author} {\bibfnamefont {S.}~\bibnamefont {Benhamou}},\
  }\bibfield  {title} {\bibinfo {title} {Random walk models in biology},\
  }\href {https://doi.org/10.1098%2Frsif.2008.0014} {\bibfield  {journal}
  {\bibinfo  {journal} {Journal of The Royal Society Interface}\ }\textbf
  {\bibinfo {volume} {5}},\ \bibinfo {pages} {813} (\bibinfo {year}
  {2008})}\BibitemShut {NoStop}%
\bibitem [{\citenamefont {Taylor}\ \emph {et~al.}(2015)\citenamefont {Taylor},
  \citenamefont {Baker},\ and\ \citenamefont {Yates}}]{taylor2015dab}%
  \BibitemOpen
  \bibfield  {author} {\bibinfo {author} {\bibfnamefont {P.}~\bibnamefont
  {Taylor}}, \bibinfo {author} {\bibfnamefont {R.}~\bibnamefont {Baker}},\ and\
  \bibinfo {author} {\bibfnamefont {C.}~\bibnamefont {Yates}},\ }\bibfield
  {title} {\bibinfo {title} {Deriving appropriate boundary conditions, and
  accelerating position-jump simulations, of diffusion using non-local
  jumping},\ }\href@noop {} {\bibfield  {journal} {\bibinfo  {journal} {Phys.
  Biol.}\ }\textbf {\bibinfo {volume} {12}},\ \bibinfo {pages} {016006}
  (\bibinfo {year} {2015})}\BibitemShut {NoStop}%
\bibitem [{\citenamefont {Dickman}(1989)}]{Dickman_1989}%
  \BibitemOpen
  \bibfield  {author} {\bibinfo {author} {\bibfnamefont {R.}~\bibnamefont
  {Dickman}},\ }\bibfield  {title} {\bibinfo {title} {Nonequilibrium lattice
  models: Series analysis of steady states},\ }\href
  {https://doi.org/10.1007%2Fbf01041076} {\bibfield  {journal} {\bibinfo
  {journal} {Journal of Statistical Physics}\ }\textbf {\bibinfo {volume}
  {55}},\ \bibinfo {pages} {997} (\bibinfo {year} {1989})}\BibitemShut
  {NoStop}%
\bibitem [{\citenamefont {Simpson}\ and\ \citenamefont
  {Baker}(2011)}]{PhysRevE.83.051922}%
  \BibitemOpen
  \bibfield  {author} {\bibinfo {author} {\bibfnamefont {M.~J.}\ \bibnamefont
  {Simpson}}\ and\ \bibinfo {author} {\bibfnamefont {R.~E.}\ \bibnamefont
  {Baker}},\ }\bibfield  {title} {\bibinfo {title} {Corrected mean-field models
  for spatially dependent advection-diffusion-reaction phenomena},\ }\href@noop
  {} {\bibfield  {journal} {\bibinfo  {journal} {Phys. Rev. E}\ }\textbf
  {\bibinfo {volume} {83}},\ \bibinfo {pages} {051922} (\bibinfo {year}
  {2011})}\BibitemShut {NoStop}%
\bibitem [{\citenamefont {Penrose}(2008)}]{Penrose_2008}%
  \BibitemOpen
  \bibfield  {author} {\bibinfo {author} {\bibfnamefont {M.~D.}\ \bibnamefont
  {Penrose}},\ }\bibfield  {title} {\bibinfo {title} {Existence and spatial
  limit theorems for lattice and continuum particle systems},\ }\href@noop {}
  {\bibfield  {journal} {\bibinfo  {journal} {Probability Surveys}\ }\textbf
  {\bibinfo {volume} {5}} (\bibinfo {year} {2008})}\BibitemShut {NoStop}%
\bibitem [{\citenamefont {Toral}\ and\ \citenamefont
  {Colet}(2014)}]{2014Toral}%
  \BibitemOpen
  \bibinfo {editor} {\bibfnamefont {R.}~\bibnamefont {Toral}}\ and\ \bibinfo
  {editor} {\bibfnamefont {P.}~\bibnamefont {Colet}},\ eds.,\ \href
  {https://doi.org/10.1002%2F9783527683147} {\emph {\bibinfo {title}
  {Stochastic Numerical Methods}}}\ (\bibinfo  {publisher} {Wiley-{VCH} Verlag
  {GmbH} {\&} Co. {KGaA}},\ \bibinfo {year} {2014})\BibitemShut {NoStop}%
\bibitem [{\citenamefont {Bick}\ \emph {et~al.}(2020)\citenamefont {Bick},
  \citenamefont {Goodfellow}, \citenamefont {Laing},\ and\ \citenamefont
  {Martens}}]{Bick_2020}%
  \BibitemOpen
  \bibfield  {author} {\bibinfo {author} {\bibfnamefont {C.}~\bibnamefont
  {Bick}}, \bibinfo {author} {\bibfnamefont {M.}~\bibnamefont {Goodfellow}},
  \bibinfo {author} {\bibfnamefont {C.~R.}\ \bibnamefont {Laing}},\ and\
  \bibinfo {author} {\bibfnamefont {E.~A.}\ \bibnamefont {Martens}},\
  }\bibfield  {title} {\bibinfo {title} {Understanding the dynamics of
  biological and neural oscillator networks through exact mean-field
  reductions: a review},\ }\href {https://doi.org/10.1186%2Fs13408-020-00086-9}
  {\bibfield  {journal} {\bibinfo  {journal} {The Journal of Mathematical
  Neuroscience}\ }\textbf {\bibinfo {volume} {10}} (\bibinfo {year}
  {2020})}\BibitemShut {NoStop}%
\bibitem [{\citenamefont {Alonso}\ \emph {et~al.}(2006)\citenamefont {Alonso},
  \citenamefont {McKane},\ and\ \citenamefont {Pascual}}]{Alonso_2006}%
  \BibitemOpen
  \bibfield  {author} {\bibinfo {author} {\bibfnamefont {D.}~\bibnamefont
  {Alonso}}, \bibinfo {author} {\bibfnamefont {A.~J.}\ \bibnamefont {McKane}},\
  and\ \bibinfo {author} {\bibfnamefont {M.}~\bibnamefont {Pascual}},\
  }\bibfield  {title} {\bibinfo {title} {Stochastic amplification in
  epidemics},\ }\href {https://doi.org/10.1098%2Frsif.2006.0192} {\bibfield
  {journal} {\bibinfo  {journal} {Journal of The Royal Society Interface}\
  }\textbf {\bibinfo {volume} {4}},\ \bibinfo {pages} {575} (\bibinfo {year}
  {2006})}\BibitemShut {NoStop}%
\bibitem [{\citenamefont {Biancalani}\ \emph {et~al.}(2014)\citenamefont
  {Biancalani}, \citenamefont {Dyson},\ and\ \citenamefont
  {McKane}}]{PhysRevLett.112.038101}%
  \BibitemOpen
  \bibfield  {author} {\bibinfo {author} {\bibfnamefont {T.}~\bibnamefont
  {Biancalani}}, \bibinfo {author} {\bibfnamefont {L.}~\bibnamefont {Dyson}},\
  and\ \bibinfo {author} {\bibfnamefont {A.~J.}\ \bibnamefont {McKane}},\
  }\bibfield  {title} {\bibinfo {title} {Noise-induced bistable states and
  their mean switching time in foraging colonies},\ }\href
  {https://link.aps.org/doi/10.1103/PhysRevLett.112.038101} {\bibfield
  {journal} {\bibinfo  {journal} {Phys. Rev. Lett.}\ }\textbf {\bibinfo
  {volume} {112}},\ \bibinfo {pages} {038101} (\bibinfo {year}
  {2014})}\BibitemShut {NoStop}%
\bibitem [{\citenamefont {Constable}\ \emph {et~al.}(2016)\citenamefont
  {Constable}, \citenamefont {Rogers}, \citenamefont {McKane},\ and\
  \citenamefont {Tarnita}}]{Constable_2016}%
  \BibitemOpen
  \bibfield  {author} {\bibinfo {author} {\bibfnamefont {G.~W.~A.}\
  \bibnamefont {Constable}}, \bibinfo {author} {\bibfnamefont {T.}~\bibnamefont
  {Rogers}}, \bibinfo {author} {\bibfnamefont {A.~J.}\ \bibnamefont {McKane}},\
  and\ \bibinfo {author} {\bibfnamefont {C.~E.}\ \bibnamefont {Tarnita}},\
  }\bibfield  {title} {\bibinfo {title} {Demographic noise can reverse the
  direction of deterministic selection},\ }\href
  {https://doi.org/10.1073%2Fpnas.1603693113} {\bibfield  {journal} {\bibinfo
  {journal} {Proceedings of the National Academy of Sciences}\ }\textbf
  {\bibinfo {volume} {113}} (\bibinfo {year} {2016})}\BibitemShut {NoStop}%
\bibitem [{\citenamefont {Funk}\ \emph {et~al.}(2009)\citenamefont {Funk},
  \citenamefont {Gilad}, \citenamefont {Watkins},\ and\ \citenamefont
  {Jansen}}]{Funk2009}%
  \BibitemOpen
  \bibfield  {author} {\bibinfo {author} {\bibfnamefont {S.}~\bibnamefont
  {Funk}}, \bibinfo {author} {\bibfnamefont {E.}~\bibnamefont {Gilad}},
  \bibinfo {author} {\bibfnamefont {C.}~\bibnamefont {Watkins}},\ and\ \bibinfo
  {author} {\bibfnamefont {V.~A.~A.}\ \bibnamefont {Jansen}},\ }\bibfield
  {title} {\bibinfo {title} {The spread of awareness and its impact on epidemic
  outbreaks},\ }\href {https://doi.org/10.1073/pnas.0810762106} {\bibfield
  {journal} {\bibinfo  {journal} {Proceedings of the National Academy of
  Sciences}\ }\textbf {\bibinfo {volume} {106}},\ \bibinfo {pages} {6872}
  (\bibinfo {year} {2009})}\BibitemShut {NoStop}%
\bibitem [{\citenamefont {Felsenstein}(1974)}]{10.1093/genetics/78.2.737}%
  \BibitemOpen
  \bibfield  {author} {\bibinfo {author} {\bibfnamefont {J.}~\bibnamefont
  {Felsenstein}},\ }\bibfield  {title} {\bibinfo {title} {{The evolutionary
  advantage of recombination}},\ }\href@noop {} {\bibfield  {journal} {\bibinfo
   {journal} {Genetics}\ }\textbf {\bibinfo {volume} {78}},\ \bibinfo {pages}
  {737} (\bibinfo {year} {1974})}\BibitemShut {NoStop}%
\bibitem [{\citenamefont {Brunet}\ and\ \citenamefont
  {Derrida}(1997)}]{Brunet_1997}%
  \BibitemOpen
  \bibfield  {author} {\bibinfo {author} {\bibfnamefont {E.}~\bibnamefont
  {Brunet}}\ and\ \bibinfo {author} {\bibfnamefont {B.}~\bibnamefont
  {Derrida}},\ }\bibfield  {title} {\bibinfo {title} {Shift in the velocity of
  a front due to a cutoff},\ }\href
  {https://doi.org/10.1103%2Fphysreve.56.2597} {\bibfield  {journal} {\bibinfo
  {journal} {Physical Review E}\ }\textbf {\bibinfo {volume} {56}},\ \bibinfo
  {pages} {2597} (\bibinfo {year} {1997})}\BibitemShut {NoStop}%
\bibitem [{\citenamefont {Baxendale}\ and\ \citenamefont
  {Greenwood}(2010)}]{Baxendale2010}%
  \BibitemOpen
  \bibfield  {author} {\bibinfo {author} {\bibfnamefont {P.~H.}\ \bibnamefont
  {Baxendale}}\ and\ \bibinfo {author} {\bibfnamefont {P.~E.}\ \bibnamefont
  {Greenwood}},\ }\bibfield  {title} {\bibinfo {title} {Sustained oscillations
  for density dependent markov processes},\ }\href@noop {} {\bibfield
  {journal} {\bibinfo  {journal} {Journal of Mathematical Biology}\ }\textbf
  {\bibinfo {volume} {63}},\ \bibinfo {pages} {433} (\bibinfo {year}
  {2010})}\BibitemShut {NoStop}%
\bibitem [{\citenamefont {{VAN KAMPEN}}(2007)}]{VANKAMPEN2007244}%
  \BibitemOpen
  \bibfield  {author} {\bibinfo {author} {\bibfnamefont {N.}~\bibnamefont {{VAN
  KAMPEN}}},\ }\bibfield  {title} {\bibinfo {title} {Chapter x - the expansion
  of the master equation},\ }in\ \href
  {https://doi.org/https://doi.org/10.1016/B978-044452965-7/50013-1} {\emph
  {\bibinfo {booktitle} {Stochastic Processes in Physics and Chemistry (Third
  Edition)}}},\ \bibinfo {series and number} {North-Holland Personal Library},\
  \bibinfo {editor} {edited by\ \bibinfo {editor} {\bibfnamefont
  {N.}~\bibnamefont {{VAN KAMPEN}}}}\ (\bibinfo  {publisher} {Elsevier},\
  \bibinfo {address} {Amsterdam},\ \bibinfo {year} {2007})\ \bibinfo {edition}
  {third edition}\ ed.,\ pp.\ \bibinfo {pages} {244--272}\BibitemShut {NoStop}%
\bibitem [{\citenamefont {Kurtz}(1978)}]{KURTZ1978223}%
  \BibitemOpen
  \bibfield  {author} {\bibinfo {author} {\bibfnamefont {T.~G.}\ \bibnamefont
  {Kurtz}},\ }\bibfield  {title} {\bibinfo {title} {Strong approximation
  theorems for density dependent markov chains},\ }\href
  {https://doi.org/https://doi.org/10.1016/0304-4149(78)90020-0} {\bibfield
  {journal} {\bibinfo  {journal} {Stochastic Processes and their Applications}\
  }\textbf {\bibinfo {volume} {6}},\ \bibinfo {pages} {223} (\bibinfo {year}
  {1978})}\BibitemShut {NoStop}%
\bibitem [{\citenamefont {Parsons}(2018)}]{Parsons2018}%
  \BibitemOpen
  \bibfield  {author} {\bibinfo {author} {\bibfnamefont {T.~L.}\ \bibnamefont
  {Parsons}},\ }\bibfield  {title} {\bibinfo {title} {Invasion probabilities,
  hitting times, and some fluctuation theory for the stochastic logistic
  process},\ }\href {https://doi.org/10.1007/s00285-018-1250-x} {\bibfield
  {journal} {\bibinfo  {journal} {Journal of Mathematical Biology}\ }\textbf
  {\bibinfo {volume} {77}},\ \bibinfo {pages} {1193} (\bibinfo {year}
  {2018})}\BibitemShut {NoStop}%
\bibitem [{\citenamefont {Dumortier}\ \emph {et~al.}(2007)\citenamefont
  {Dumortier}, \citenamefont {Popovi{\'{c}}},\ and\ \citenamefont
  {Kaper}}]{Dumortier_2007}%
  \BibitemOpen
  \bibfield  {author} {\bibinfo {author} {\bibfnamefont {F.}~\bibnamefont
  {Dumortier}}, \bibinfo {author} {\bibfnamefont {N.}~\bibnamefont
  {Popovi{\'{c}}}},\ and\ \bibinfo {author} {\bibfnamefont {T.~J.}\
  \bibnamefont {Kaper}},\ }\bibfield  {title} {\bibinfo {title} {{The critical
  wave speed for the
  Fisher{\textendash}Kolmogorov{\textendash}Petrowskii{\textendash}Piscounov
  equation with cut-off}},\ }\href@noop {} {\bibfield  {journal} {\bibinfo
  {journal} {Nonlinearity}\ }\textbf {\bibinfo {volume} {20}},\ \bibinfo
  {pages} {855} (\bibinfo {year} {2007})}\BibitemShut {NoStop}%
\bibitem [{\citenamefont {Dumortier}\ and\ \citenamefont
  {Kaper}(2014)}]{Dumortier_2014}%
  \BibitemOpen
  \bibfield  {author} {\bibinfo {author} {\bibfnamefont {F.}~\bibnamefont
  {Dumortier}}\ and\ \bibinfo {author} {\bibfnamefont {T.~J.}\ \bibnamefont
  {Kaper}},\ }\bibfield  {title} {\bibinfo {title} {Wave speeds for the {FKPP}
  equation with enhancements of the reaction function},\ }\href
  {https://doi.org/10.1007%2Fs00033-014-0422-9} {\bibfield  {journal} {\bibinfo
   {journal} {Zeitschrift für angewandte Mathematik und Physik}\ }\textbf
  {\bibinfo {volume} {66}},\ \bibinfo {pages} {607} (\bibinfo {year}
  {2014})}\BibitemShut {NoStop}%
\bibitem [{\citenamefont {Brunet}\ \emph {et~al.}(2006)\citenamefont {Brunet},
  \citenamefont {Derrida}, \citenamefont {Mueller},\ and\ \citenamefont
  {Munier}}]{Brunet_2006}%
  \BibitemOpen
  \bibfield  {author} {\bibinfo {author} {\bibfnamefont {E.}~\bibnamefont
  {Brunet}}, \bibinfo {author} {\bibfnamefont {B.}~\bibnamefont {Derrida}},
  \bibinfo {author} {\bibfnamefont {A.~H.}\ \bibnamefont {Mueller}},\ and\
  \bibinfo {author} {\bibfnamefont {S.}~\bibnamefont {Munier}},\ }\bibfield
  {title} {\bibinfo {title} {Noisy traveling waves: Effect of selection on
  genealogies},\ }\href {https://doi.org/10.1209%2Fepl%2Fi2006-10224-4}
  {\bibfield  {journal} {\bibinfo  {journal} {Europhysics Letters ({EPL})}\
  }\textbf {\bibinfo {volume} {76}},\ \bibinfo {pages} {1} (\bibinfo {year}
  {2006})}\BibitemShut {NoStop}%
\bibitem [{\citenamefont {B{\'{e}}rard}\ and\ \citenamefont
  {Gou{\'{e}}r{\'{e}}}(2010)}]{Brard2010}%
  \BibitemOpen
  \bibfield  {author} {\bibinfo {author} {\bibfnamefont {J.}~\bibnamefont
  {B{\'{e}}rard}}\ and\ \bibinfo {author} {\bibfnamefont {J.-B.}\ \bibnamefont
  {Gou{\'{e}}r{\'{e}}}},\ }\bibfield  {title} {\bibinfo {title} {Brunet-derrida
  behavior of branching-selection particle systems on the line},\ }\href@noop
  {} {\bibfield  {journal} {\bibinfo  {journal} {Communications in Mathematical
  Physics}\ }\textbf {\bibinfo {volume} {298}},\ \bibinfo {pages} {323}
  (\bibinfo {year} {2010})}\BibitemShut {NoStop}%
\bibitem [{\citenamefont {Breuer}\ \emph {et~al.}(1994)\citenamefont {Breuer},
  \citenamefont {Huber},\ and\ \citenamefont {Petruccione}}]{BREUER1994259}%
  \BibitemOpen
  \bibfield  {author} {\bibinfo {author} {\bibfnamefont {H.}~\bibnamefont
  {Breuer}}, \bibinfo {author} {\bibfnamefont {W.}~\bibnamefont {Huber}},\ and\
  \bibinfo {author} {\bibfnamefont {F.}~\bibnamefont {Petruccione}},\
  }\bibfield  {title} {\bibinfo {title} {Fluctuation effects on wave
  propagation in a reaction-diffusion process},\ }\href
  {https://www.sciencedirect.com/science/article/pii/0167278994901619}
  {\bibfield  {journal} {\bibinfo  {journal} {Physica D: Nonlinear Phenomena}\
  }\textbf {\bibinfo {volume} {73}},\ \bibinfo {pages} {259} (\bibinfo {year}
  {1994})}\BibitemShut {NoStop}%
\bibitem [{\citenamefont {Kessler}\ \emph {et~al.}(1998)\citenamefont
  {Kessler}, \citenamefont {Ner},\ and\ \citenamefont {Sander}}]{Kessler1998}%
  \BibitemOpen
  \bibfield  {author} {\bibinfo {author} {\bibfnamefont {D.~A.}\ \bibnamefont
  {Kessler}}, \bibinfo {author} {\bibfnamefont {Z.}~\bibnamefont {Ner}},\ and\
  \bibinfo {author} {\bibfnamefont {L.~M.}\ \bibnamefont {Sander}},\ }\bibfield
   {title} {\bibinfo {title} {Front propagation: Precursors, cutoffs, and
  structural stability},\ }\href@noop {} {\bibfield  {journal} {\bibinfo
  {journal} {Physical Review E}\ }\textbf {\bibinfo {volume} {58}},\ \bibinfo
  {pages} {107} (\bibinfo {year} {1998})}\BibitemShut {NoStop}%
\bibitem [{\citenamefont {Panja}\ and\ \citenamefont {van
  Saarloos}(2002{\natexlab{a}})}]{Panja_2002}%
  \BibitemOpen
  \bibfield  {author} {\bibinfo {author} {\bibfnamefont {D.}~\bibnamefont
  {Panja}}\ and\ \bibinfo {author} {\bibfnamefont {W.}~\bibnamefont {van
  Saarloos}},\ }\bibfield  {title} {\bibinfo {title} {Fronts with a growth
  cutoff but with speed higher than the linear spreading speed},\ }\href
  {https://doi.org/10.1103%2Fphysreve.66.015206} {\bibfield  {journal}
  {\bibinfo  {journal} {Physical Review E}\ }\textbf {\bibinfo {volume} {66}}
  (\bibinfo {year} {2002}{\natexlab{a}})}\BibitemShut {NoStop}%
\bibitem [{\citenamefont {Pechenik}\ and\ \citenamefont
  {Levine}(1999)}]{PhysRevE.59.3893}%
  \BibitemOpen
  \bibfield  {author} {\bibinfo {author} {\bibfnamefont {L.}~\bibnamefont
  {Pechenik}}\ and\ \bibinfo {author} {\bibfnamefont {H.}~\bibnamefont
  {Levine}},\ }\bibfield  {title} {\bibinfo {title} {Interfacial velocity
  corrections due to multiplicative noise},\ }\href@noop {} {\bibfield
  {journal} {\bibinfo  {journal} {Phys. Rev. E}\ }\textbf {\bibinfo {volume}
  {59}},\ \bibinfo {pages} {3893} (\bibinfo {year} {1999})}\BibitemShut
  {NoStop}%
\bibitem [{\citenamefont {Kolmogorov}\ \emph {et~al.}(1937)\citenamefont
  {Kolmogorov}, \citenamefont {Petrovsky},\ and\ \citenamefont
  {Piscounov}}]{kpp_equation}%
  \BibitemOpen
  \bibfield  {author} {\bibinfo {author} {\bibfnamefont {A.}~\bibnamefont
  {Kolmogorov}}, \bibinfo {author} {\bibfnamefont {I.}~\bibnamefont
  {Petrovsky}},\ and\ \bibinfo {author} {\bibfnamefont {N.}~\bibnamefont
  {Piscounov}},\ }\bibfield  {title} {\bibinfo {title} {Etude de l'equation de
  la diffusion avec croissance de la quantite de matiere et son application a
  un probleme biologique},\ }\href@noop {} {\bibfield  {journal} {\bibinfo
  {journal} {Bull. Univ. Moskow, Ser. Internat., Sec. A}\ }\textbf {\bibinfo
  {volume} {1}},\ \bibinfo {pages} {1} (\bibinfo {year} {1937})}\BibitemShut
  {NoStop}%
\bibitem [{\citenamefont {Sontag}\ \emph {et~al.}(2022)\citenamefont {Sontag},
  \citenamefont {Rogers},\ and\ \citenamefont {Yates}}]{Sontag_2022}%
  \BibitemOpen
  \bibfield  {author} {\bibinfo {author} {\bibfnamefont {A.}~\bibnamefont
  {Sontag}}, \bibinfo {author} {\bibfnamefont {T.}~\bibnamefont {Rogers}},\
  and\ \bibinfo {author} {\bibfnamefont {C.~A.}\ \bibnamefont {Yates}},\
  }\bibfield  {title} {\bibinfo {title} {Misinformation can prevent the
  suppression of epidemics},\ }\href {https://doi.org/10.1098%2Frsif.2021.0668}
  {\bibfield  {journal} {\bibinfo  {journal} {Journal of The Royal Society
  Interface}\ }\textbf {\bibinfo {volume} {19}} (\bibinfo {year}
  {2022})}\BibitemShut {NoStop}%
\bibitem [{\citenamefont {Rogers}\ \emph {et~al.}(2012)\citenamefont {Rogers},
  \citenamefont {McKane},\ and\ \citenamefont {Rossberg}}]{Rogers_2012}%
  \BibitemOpen
  \bibfield  {author} {\bibinfo {author} {\bibfnamefont {T.}~\bibnamefont
  {Rogers}}, \bibinfo {author} {\bibfnamefont {A.~J.}\ \bibnamefont {McKane}},\
  and\ \bibinfo {author} {\bibfnamefont {A.~G.}\ \bibnamefont {Rossberg}},\
  }\bibfield  {title} {\bibinfo {title} {Demographic noise can lead to the
  spontaneous formation of species},\ }\href@noop {} {\bibfield  {journal}
  {\bibinfo  {journal} {{EPL} (Europhysics Letters)}\ }\textbf {\bibinfo
  {volume} {97}},\ \bibinfo {pages} {40008} (\bibinfo {year}
  {2012})}\BibitemShut {NoStop}%
\bibitem [{\citenamefont {Fuentes}\ \emph {et~al.}(2003)\citenamefont
  {Fuentes}, \citenamefont {Kuperman},\ and\ \citenamefont
  {Kenkre}}]{Fuentes_2003}%
  \BibitemOpen
  \bibfield  {author} {\bibinfo {author} {\bibfnamefont {M.~A.}\ \bibnamefont
  {Fuentes}}, \bibinfo {author} {\bibfnamefont {M.~N.}\ \bibnamefont
  {Kuperman}},\ and\ \bibinfo {author} {\bibfnamefont {V.~M.}\ \bibnamefont
  {Kenkre}},\ }\bibfield  {title} {\bibinfo {title} {Nonlocal interaction
  effects on pattern formation in population dynamics},\ }\href@noop {}
  {\bibfield  {journal} {\bibinfo  {journal} {Physical Review Letters}\
  }\textbf {\bibinfo {volume} {91}} (\bibinfo {year} {2003})}\BibitemShut
  {NoStop}%
\bibitem [{\citenamefont {Stevens}\ and\ \citenamefont
  {Othmer}(1997)}]{Stevens_1997}%
  \BibitemOpen
  \bibfield  {author} {\bibinfo {author} {\bibfnamefont {A.}~\bibnamefont
  {Stevens}}\ and\ \bibinfo {author} {\bibfnamefont {H.~G.}\ \bibnamefont
  {Othmer}},\ }\bibfield  {title} {\bibinfo {title} {Aggregation, blowup, and
  collapse: The {ABC}{\textquotesingle}s of taxis in reinforced random walks},\
  }\href@noop {} {\bibfield  {journal} {\bibinfo  {journal} {{SIAM} Journal on
  Applied Mathematics}\ }\textbf {\bibinfo {volume} {57}},\ \bibinfo {pages}
  {1044} (\bibinfo {year} {1997})}\BibitemShut {NoStop}%
\bibitem [{\citenamefont {Baker}\ \emph {et~al.}(2010)\citenamefont {Baker},
  \citenamefont {Yates},\ and\ \citenamefont {Erban}}]{baker2009fmm}%
  \BibitemOpen
  \bibfield  {author} {\bibinfo {author} {\bibfnamefont {R.}~\bibnamefont
  {Baker}}, \bibinfo {author} {\bibfnamefont {C.}~\bibnamefont {Yates}},\ and\
  \bibinfo {author} {\bibfnamefont {R.}~\bibnamefont {Erban}},\ }\bibfield
  {title} {\bibinfo {title} {{From microscopic to macroscopic descriptions of
  cell migration on growing domains}},\ }\href@noop {} {\bibfield  {journal}
  {\bibinfo  {journal} {Bull. Math. Biol.}\ }\textbf {\bibinfo {volume} {72}},\
  \bibinfo {pages} {719} (\bibinfo {year} {2010})}\BibitemShut {NoStop}%
\bibitem [{\citenamefont {Koll{\'{a}}r}\ and\ \citenamefont
  {Novak}(2016)}]{Kollr2016}%
  \BibitemOpen
  \bibfield  {author} {\bibinfo {author} {\bibfnamefont {R.}~\bibnamefont
  {Koll{\'{a}}r}}\ and\ \bibinfo {author} {\bibfnamefont {S.}~\bibnamefont
  {Novak}},\ }\bibfield  {title} {\bibinfo {title} {Existence of traveling
  waves for the generalized f{\textendash}{KPP} equation},\ }\href@noop {}
  {\bibfield  {journal} {\bibinfo  {journal} {Bulletin of Mathematical
  Biology}\ }\textbf {\bibinfo {volume} {79}},\ \bibinfo {pages} {525}
  (\bibinfo {year} {2016})}\BibitemShut {NoStop}%
\bibitem [{\citenamefont {Murray}(2002)}]{2002Murray}%
  \BibitemOpen
  \bibinfo {editor} {\bibfnamefont {J.~D.}\ \bibnamefont {Murray}},\ ed.,\
  \href {https://doi.org/10.1007%2Fb98868} {\emph {\bibinfo {title}
  {Mathematical Biology}}}\ (\bibinfo  {publisher} {Springer New York},\
  \bibinfo {year} {2002})\BibitemShut {NoStop}%
\bibitem [{\citenamefont {Panja}\ and\ \citenamefont {van
  Saarloos}(2002{\natexlab{b}})}]{PhysRevE.66.036206}%
  \BibitemOpen
  \bibfield  {author} {\bibinfo {author} {\bibfnamefont {D.}~\bibnamefont
  {Panja}}\ and\ \bibinfo {author} {\bibfnamefont {W.}~\bibnamefont {van
  Saarloos}},\ }\bibfield  {title} {\bibinfo {title} {{Fluctuating pulled
  fronts: The origin and the effects of a finite particle cutoff}},\ }\href
  {https://link.aps.org/doi/10.1103/PhysRevE.66.036206} {\bibfield  {journal}
  {\bibinfo  {journal} {Phys. Rev. E}\ }\textbf {\bibinfo {volume} {66}},\
  \bibinfo {pages} {036206} (\bibinfo {year} {2002}{\natexlab{b}})}\BibitemShut
  {NoStop}%
\bibitem [{\citenamefont {Holmes}\ \emph {et~al.}(1994)\citenamefont {Holmes},
  \citenamefont {Lewis}, \citenamefont {Banks},\ and\ \citenamefont
  {Veit}}]{Holmes_1994}%
  \BibitemOpen
  \bibfield  {author} {\bibinfo {author} {\bibfnamefont {E.~E.}\ \bibnamefont
  {Holmes}}, \bibinfo {author} {\bibfnamefont {M.~A.}\ \bibnamefont {Lewis}},
  \bibinfo {author} {\bibfnamefont {J.~E.}\ \bibnamefont {Banks}},\ and\
  \bibinfo {author} {\bibfnamefont {R.~R.}\ \bibnamefont {Veit}},\ }\bibfield
  {title} {\bibinfo {title} {Partial differential equations in ecology: Spatial
  interactions and population dynamics},\ }\href
  {https://doi.org/10.2307%2F1939378} {\bibfield  {journal} {\bibinfo
  {journal} {Ecology}\ }\textbf {\bibinfo {volume} {75}},\ \bibinfo {pages}
  {17} (\bibinfo {year} {1994})}\BibitemShut {NoStop}%
\bibitem [{\citenamefont {Lemarchand}\ and\ \citenamefont
  {Nowakowski}(1999)}]{doi:10.1063/1.479923}%
  \BibitemOpen
  \bibfield  {author} {\bibinfo {author} {\bibfnamefont {A.}~\bibnamefont
  {Lemarchand}}\ and\ \bibinfo {author} {\bibfnamefont {B.}~\bibnamefont
  {Nowakowski}},\ }\bibfield  {title} {\bibinfo {title} {Different description
  levels of chemical wave front and propagation speed selection},\ }\href
  {https://doi.org/10.1063/1.479923} {\bibfield  {journal} {\bibinfo  {journal}
  {The Journal of Chemical Physics}\ }\textbf {\bibinfo {volume} {111}},\
  \bibinfo {pages} {6190} (\bibinfo {year} {1999})}\BibitemShut {NoStop}%
\bibitem [{\citenamefont {Panja}(2004)}]{PANJA200487}%
  \BibitemOpen
  \bibfield  {author} {\bibinfo {author} {\bibfnamefont {D.}~\bibnamefont
  {Panja}},\ }\bibfield  {title} {\bibinfo {title} {Effects of fluctuations on
  propagating fronts},\ }\href
  {https://www.sciencedirect.com/science/article/pii/S0370157303004630}
  {\bibfield  {journal} {\bibinfo  {journal} {Physics Reports}\ }\textbf
  {\bibinfo {volume} {393}},\ \bibinfo {pages} {87} (\bibinfo {year}
  {2004})}\BibitemShut {NoStop}%
\bibitem [{\citenamefont {Moro}(2001)}]{PhysRevLett.87.238303}%
  \BibitemOpen
  \bibfield  {author} {\bibinfo {author} {\bibfnamefont {E.}~\bibnamefont
  {Moro}},\ }\bibfield  {title} {\bibinfo {title} {Internal fluctuations
  effects on fisher waves},\ }\href@noop {} {\bibfield  {journal} {\bibinfo
  {journal} {Phys. Rev. Lett.}\ }\textbf {\bibinfo {volume} {87}},\ \bibinfo
  {pages} {238303} (\bibinfo {year} {2001})}\BibitemShut {NoStop}%
\bibitem [{\citenamefont {Panja}\ and\ \citenamefont {van
  Saarloos}(2002{\natexlab{c}})}]{PhysRevE.65.057202}%
  \BibitemOpen
  \bibfield  {author} {\bibinfo {author} {\bibfnamefont {D.}~\bibnamefont
  {Panja}}\ and\ \bibinfo {author} {\bibfnamefont {W.}~\bibnamefont {van
  Saarloos}},\ }\bibfield  {title} {\bibinfo {title} {Weakly pushed nature of
  ``pulled'' fronts with a cutoff},\ }\href@noop {} {\bibfield  {journal}
  {\bibinfo  {journal} {Phys. Rev. E}\ }\textbf {\bibinfo {volume} {65}},\
  \bibinfo {pages} {057202} (\bibinfo {year} {2002}{\natexlab{c}})}\BibitemShut
  {NoStop}%
\bibitem [{\citenamefont {Panja}(2003)}]{PhysRevE.68.065202}%
  \BibitemOpen
  \bibfield  {author} {\bibinfo {author} {\bibfnamefont {D.}~\bibnamefont
  {Panja}},\ }\bibfield  {title} {\bibinfo {title} {Asymptotic scaling of the
  diffusion coefficient of fluctuating ``pulled'' fronts},\ }\href@noop {}
  {\bibfield  {journal} {\bibinfo  {journal} {Phys. Rev. E}\ }\textbf {\bibinfo
  {volume} {68}},\ \bibinfo {pages} {065202} (\bibinfo {year}
  {2003})}\BibitemShut {NoStop}%
\end{thebibliography}%
\end{document}